\documentclass[preprint]{aastex62}
\submitjournal{ApJ}

\shorttitle{Regime transition of general circulation on High Obliquity Planets}
\shortauthors{Kang}
\begin{document}

\title{Regime transition between eddy-driven and moist-driven circulation on High Obliquity Planets}

\correspondingauthor{Wanying Kang}
\email{wanyingkang@g.harvard.edu}

\author[0000-0002-4615-3702]{Wanying Kang}
\affil{School of Engineering and Applied Sciences\\
  Harvard University\\
  Cambridge, MA 02138, USA}

% \begin{keypoints}
% \item 
% \item 
% \item 
% \end{keypoints}

\begin{abstract}
    We investigate how the meridional circulation and baroclinic eddies change with insolation and rotation rate, under high and zero obliquity setups, using a general circulation model. The total circulation is considered as superposition of circulations driven by different physics processes, such as diabatic and adiabatic processes. We decompose the meridional circulation into diabatic and adiabatic components, in order to understand their different responses to changes of insolation and rotation rate.
  
  As insolation or rotation period increases, the meridional circulation tends to become more diabatically dominant, regardless of the obliquity. The low obliquity circulation is always dominated by diabatic processes, while the high obliquity configuration has two circulation regimes: an adiabatic-dominant regime in the limit of low insolation and fast rotation, and a diabatic-dominant regime in the opposite limit. This regime transition may be observable via its signature on the upper atmospheric zonal wind and the column cloud cover. The momentum-driven circulation, the dominant circulation component in the weak-insolation and fast-rotating regimes is found to resemble that in a dry dynamic model forced by a reversed meridional temperature gradient, indicating the relevance of using a dry dynamic model to understand planetary general circulations under high obliquity.
\end{abstract}

%% Keywords should appear after the \end{abstract} command. 
%% See the online documentation for the full list of available subject
%% keywords and the rules for their use.

\section{Introduction}
\label{sec:introduction}

%% Importance of general circulation in general
While single column radiative convective equilibrium (RCE) models capture many dominant features of a climate system, it ignored the horizontal inhomogeneity, an important factor involved in most climate feedbacks. For example, the activation of ice-albedo feedback is controlled by the lowest rather than the average surface temperature over the globe, and the former is to a large extent affected by the temporal and spatial distribution of insolation, and the horizontal heat transport \citep[e.g,][]{Yang-Boue-Fabrycky-et-al-2014:strong, Linsenmeier-Pascale-Lucarini-2015:climate}. Cloud cover and its radiative effects could change completely when the general circulation changes \citep{Wang-Liu-Tian-et-al-2016:effects, Kang-2019:mechanisms}. Therefore, 3D general circulation models (GCM) are necessary in order to account for these feedbacks correctly.

% 0.99-1.70 AU for an Earth-like planet in 1D RCM  \citep{Kopparapu-Ramirez-Kasting-et-al-2013:erratum}
% and when accounting for cloud feedbacks \citep{Leconte-Forget-Charnay-et-al-2013:climate, Yang-Cowan-Abbot-2013:stabilizing}.

%% Reason for study high obliquity: exoplanet
High obliquity planets are expected to widely exist in the universe. In our solar system, for example, Mars has an obliquity chaotically varying from 0 to 60 degree \citep{Laskar-Robutel-1993:chaotic}, and Venus and Uranus have obliquities close to 180 and 90 degree respectively \citep{Carpenter-1966:study}.
Exoplanets may have a large obliquity or large obliquity variance if they are not very close to the host star and if they do not have a large moon \citep{Heller-Leconte-Barnes-2011:tidal}, depending on the initial angular momentum of the nebulae that formed the planet, continental movement \citep{Williams-Kasting-Frakes-1998:low}, gravitational interaction with other bodies \citep{Correia-Laskar-2010:tidal}, the history of orbital migration \citep{Brunini-2006:origin} and the secular resonance-driven spin-orbit coupling \citep{Millholland-Laughlin-2019:obliquity}. 

%% obliquity makes a difference.
Obliquity is one of the orbital factors that are missed in the single column picture. Although obliquity has no effect on the global annual mean insolation, it may have contributed to setting the timing of the transitions between glacial and inter-glacial states during Pleistocene \citep{Huybers-Wunsch-2005:obliquity, Schulz-Zeebe-2006:pleistocene}, especially via nonlinear phase locking \citep{Tziperman-Raymo-Huybers-et-al-2006:consequences}.
Under high obliquity, exoplanets are found in GCMs to remain habitable under much lower insolation than their low obliquity equivalents \citep{Linsenmeier-Pascale-Lucarini-2015:climate, Kilic-Raible-Stocker-2017:multiple, Kilic-Lunkeit-Raible-et-al-2018:stable, Rose-Cronin-Bitz-2017:ice, Armstrong-Barnes-Domagal-Goldman-et-al-2014:effects}, because permanent ice tends not to form under high obliquity, \citep{Linsenmeier-Pascale-Lucarini-2015:climate, Kilic-Lunkeit-Raible-et-al-2018:stable}, giving rise to lower planetary albedo. The moist greenhouse state \citep{Kasting-Pollack-1983:loss} is more likely to occur under high obliquity, due to the strong seasonal cycle and the changes of the stratospheric circulation, ending the habitable zone before the runaway greenhouse kicks in \citep{Kang-2019:wetter}. Within the habitable zone, high obliquity states tend to be warmer than low obliquity states with all other parameters fixed \citep{Jenkins-2003:gcm}, because ice cover is low \citep{Linsenmeier-Pascale-Lucarini-2015:climate, Kilic-Lunkeit-Raible-et-al-2018:stable} and because clouds (surface temperature) lag behind the substellar point due to the surface heat inertia, both leading to ineffective sunlight reflection \citep{Kang-2019:mechanisms}.

%% Previous studies on meridional circulation, including our reverse gradient study
Meridional circulation and baroclinic eddies are crucial to the understanding of climate. They carry heat, shaping the surface temperature distribution, while the vertical motions control cloud formation, affecting the planetary albedo. Clouds may be observable \citep[e.g.,][]{Demory-Wit-Lewis-et-al-2013:inference}. Our purpose here is to understand how the meridional circulation and baroclinic eddies respond to changes in insolation, rotation rate and obliquity, and to find out potential ways to observe such changes. As far as we know, we are the first to investigate the response of the high-obliquity meridional circulation to orbital parameters, and also the first to understand the response through circulation decomposition. 

The atmospheric general circulation under low obliquity has been thoroughly studied in the Earth's context. While for high obliquity planets, \citet{Linsenmeier-Pascale-Lucarini-2015:climate, Kilic-Raible-Stocker-2017:multiple} and \citet{Kilic-Lunkeit-Raible-et-al-2018:stable} simulated the atmospheric meridional circulation in both annual mean and solstice season, but the circulation pattern is yet to be understood and how the circulation changes with other orbital parameters is yet to be explored. \citet[][hereafter KCT]{Kang-Cai-Tziperman-2019:tropical} reversed the meridional temperature gradient in a dry dynamic core model, as inspired by \citet{Ferreira-Marshall-OGorman-et-al-2014:climate}, to capture the key feature of high obliquity climate, and found that the meridional circulation under high obliquity becomes weak, bottom-amplified, and found a thermally-indirect Hadley after taking annual average. These features can be deduced from the fact that the baroclinic structure is bottom amplified, as explained by a 1D linear baroclinic instability model (KCT). However, the relevance of such an idealized model study to the real high obliquity world has not been shown, and that is another main objective of this work. 

%% In this work, we will ...
In this study, we gradually change the insolation and rotation rate in a GCM, in order to study how the meridional circulation and baroclinic eddies respond, and to find potential observables.
%% section overview
The paper is organized in the following way. In section~\ref{sec:methods}, we introduce the 3D GCM setup, as well as a method used to decompose the meridional circulation into contributions by different physics processes. Results are presented in section~\ref{sec:results}. We first give an overview of the two-regime behavior (section~\ref{sec:two-regime-behavior}), then understand the high obliquity circulation pattern by connecting our finding to the previous dry model study KCT (section~\ref{sec:insights-from-dry}), and finally provide some potential observables (section~\ref{sec:obs-evidence-regime-trans}). In section~\ref{sec:conclusions}, we summarize our results.

\section{Methods}
\label{sec:methods}

\subsection{Models}
% the model in general
The model used here is modified by \citet[][code are available on GitHub\footnote{ https://github.com/storyofthewolf/ExoRT and https://github.com/storyofthewolf/ExoCAM}]{Kopparapu-Wolf-Arney-et-al-2017:habitable} based on the Community Earth System Model version 1.2.1 \citep[CESM,][]{Neale-Chen-Gettelman-2010:description}, to include mainly the following two features: 1) more realistic radiation calculation resulting from increased spectral resolution, updated spectral coefficients based on the HiTran 2012 database \citep{Rothman-Gordon-Babikov-et-al-2013:hitran2012}, and a new continuum opacity model \citep{Paynter-Ramaswamy-2014:investigating}, and 2) more frequent sub-step dynamic adjustment to improve numerical stability. The radiation transfer model does not include the CO$_2$ absorption, and thus we consider here an atmosphere made of 1 bar N$_2$ and condensible water vapor. Thanks to the fine spectral resolution, this radiation scheme was shown to be more robust at the high temperature end and to remain stable with global mean surface temperatures beyond 360 K \citep{Kopparapu-Wolf-Arney-et-al-2017:habitable}, while the default CESM radiative transfer model underestimates both longwave and shortwave water vapor absorption \citep{Yang-Leconte-Wolf-et-al-2016:differences}. Other parameters are set to the values on Earth unless otherwise mentioned.
% This advantage is not necessary to study the relative warmness of high obliquity planets under a regular insolation, but is crucial for the estimation of the inner edge habitable zone under low and high obliquity scenarios, which will be shown in the part II of this work.
The atmospheric circulation is simulated by a finite-volume dynamic core, with approximately 1.9 degree horizontal resolution and 40 vertical layers extending to 0.8 mb. This atmosphere model is coupled with a 50 meter deep slab ocean. Horizontal ocean heat transport is ignored for simplicity, and it was also shown in \citet{Jenkins-2003:gcm} to play a minor role in the surface temperature, compared to the role of the large changes in obliquity. Sea ice is simulated using Community Ice CodE (CICE) version 4, part of CESM 1.2.1. 

We gradually increase the insolation under 0 deg obliquity setup (OBL0-S) and under 80 deg obliquity setup (OBL80-S) respectively. We attempt to cover the entire habitable range, with the lowest insolation corresponding to an almost snowball state and the highest insolation corresponding to an almost run-away greenhouse state. For zero obliquity experiments, we vary the insolation from 1360 W/m$^2$ to 1760 W/m$^2$ in 80 years; and for high obliquity, from 1200 W/m$^2$ to 1750 W/m$^2$ in 110 years. There are more than one equilibrium climate state for part of the insolation range we use here due to the positive ice-albedo feedback. By increasing the insolation instead of decreasing it, we here only explore the colder branch. Similar behavior is expected for the warm branch.

We also investigate the effects of the rotation rate, by exploring the rotation rate from 5 times to 0.2 times the Earth's rotation rate. These experiments will be referred to as OBL0-$\Omega$ and OBL80-$\Omega$, with the number following ``OBL'' to be the obliquity used in the simulation. Insolation is chosen to be 1550 W/m$^2$ and 1360 W/m$^2$ for OBL80-$\Omega$ and OBL0-$\Omega$, respectively, in order to maximize the probability of identifying a regime transition (It turns out that the regime transition occurs at these insolation levels in OBL0-S and OBL80-S. Reason for these choices will be clarified later.). We initialize the model using the restart files from OBL0-S/OBL80-S simulations when the insolation reaches these levels, and increase and decrease the rotation rate to 5 times and 0.2 times earth rotation rate in 40 years, in order to keep the climate on the same branch as in OBL0-S and OBL80-S.

The memory time scale of a slab ocean simulation can be as high as 5-10 years due to the high heat content in the ocean. To demonstrate that the transient experiments almost reach the equilibrium state, we increase the insolation twice as fast, and find the evolution of the surface temperature does not change significantly for most of the regime we explored (Fig.~\ref{fig:transient-equilibrium}a). The largest disagreement shows up toward the end of the transient simulations. In particular, the high obliquity global annual mean surface temperature is about 20~K warmer in the slow transient simulations compared to the fast one. We then run two steady-state simulations at 1750~W/m$^2$ under low and high obliquity respectively, and compare the steady-state temperature field (Fig.~\ref{fig:transient-equilibrium}e-g) with that from the slow transient simulations (Fig.~\ref{fig:transient-equilibrium}b-d). The similarity indicates that the slow transient simulations are almost in equilibrium state.

We will also briefly refer to results from KCT using a dry dynamic core model. This model is a Held-Suarez type model \citep{Held-Suarez-1994:proposal}, where the 3D Navier-Stocks equation is integrated with temperature restored to a prescribed equilibrium state. In KCT, we ran the dry model with the default setup as in \citet{Held-Suarez-1994:proposal}, and with a reversed the meridional temperature gradient (the poles are warmer than the equator). The case with a reversed temperature gradient was used to study the basic dynamics of high obliquity climate, as high obliquity planets receive more radiation over the pole than the equator on annual average.

\subsection{Decomposing the meridional circulation.}
\label{sec:decompose-meridional-circulation}
%XX maybe I should throw the whole thing to supplementary.
The meridional circulation dominantly driven by adiabatic processes is different from that driven by diabatic processes, and may have different observable consequences. Thus we try to decompose the meridional circulation and understand the two regimes. We reconstruct the circulation associated with diabatic/adiabatic processes by solving the Kuo-Eliassen equation \citep{Kuo-1956:forced, Chang-1996:mean}.

The derivation of Kuo-Eliassen equation is sketched here for reference. Starting with the temperature and zonal momentum equation averaged over $(t,t+\delta t)$,
\begin{eqnarray}
  &~&\frac{u(t+\delta t)-u(t)}{\delta t}-f\bar{v} = -\frac{\bar{v}}{a\cos\theta}\partial_\theta (\bar{u}\cos\theta) -\Omega \partial_p \bar{u}-\frac{1}{a\cos^2\theta}\partial_\theta(\overline{u'v'}\cos^2\theta)-\partial_p(\overline{u'\omega'})+F \label{eq:U-equation}\\
&~&\frac{T(t+\delta t)-T(t)}{\delta t} - S_p\bar{\omega} = -\frac{\bar{v}}{a}\partial_\theta \bar{T} -\frac{1}{a\cos\theta}\partial_\theta(\overline{v'T'}\cos\theta)-\gamma\partial_p(\overline{\gamma^{-1}\omega'T'})+\frac{Q}{C_p}, \label{eq:T-equation}
\end{eqnarray}
where $F$ is the friction term, $Q$ is the diabatic heating, $S_p=-\gamma\frac{\partial \bar{\Theta}}{\partial p}$ is the static stability, $\bar{\Theta}=\bar{T}/\gamma$, and $\gamma=(p/p_0)^\kappa$. $\overline{(\cdot)}$ denotes seasonal climatology (the time derivative terms may not vanish with a seasonal cycle), and prime denotes deviations. Time average transfers the time derivative into the changes during the time interval divided by the time interval (the first term in each equation). They are not necessarily small when the time interval is a particular season. We cancel the two these time derivative terms using the thermal wind balance, $pf \bar{u}_p= R/a \bar{T}_\theta$, and represent $\bar{v}$ and $\bar{\omega}$ with the meridional streamfunction defined below, to find
\begin{eqnarray}
f^2p\frac{\partial^2\psi}{\partial p^2}&+&\frac{R}{a^2}\cos\theta\frac{\partial}{\partial\theta}\left(\frac{S_p}{\cos\theta}\frac{\partial\psi}{\partial\theta}\right)=\nonumber\\
  &-&\frac{f^2p}{a\cos^2\theta}\frac{\partial^2\left(\overline{u'v'}\cos^2\theta\right)}{\partial\theta\partial p}-f^2p\cos\theta\frac{\partial^2\overline{u'\omega'}}{\partial p^2}\nonumber\\
  &-&\frac{\cos\theta}{a^2}\frac{\partial}{\partial_\theta}\left(\frac{1}{\cos\theta}\frac{\partial\left(\overline{v'T'}\cos\theta\right)}{\partial\theta}\right) -\frac{\gamma R\cos\theta}{a}\frac{\partial^2\left(\overline{\omega'\theta'}\right)}{\partial\theta\partial p}\nonumber\\
                                                                                                                                                                    &+&f^2p\cos\theta\frac{\partial F}{\partial p}+\frac{\gamma R}{C_p}\frac{\cos\theta}{a}\frac{\partial Q}{\partial \theta} + \mathsf{advective\ terms}  \label{eq:hadley-drive}\\
  &~&\mathsf{advective\ terms}= -\frac{f^2p}{a}\frac{\partial}{\partial p}\left(\bar{v}\frac{\partial \bar{u}\cos\theta}{\partial\theta}\right)-f^2p\cos\theta\frac{\partial}{\partial p}\left(\bar{\omega} \bar{u}_p\right) -\frac{\cos\theta}{a^2}\frac{\partial}{\partial \theta}\left(\bar{v} \bar{T}_\theta\right) \label{eq:advection}
\end{eqnarray}
% \begin{eqnarray}
% &fpM_y\frac{\partial^2\psi}{\partial p^2}+\frac{RS_p}{a^2}\cos\theta\frac{\partial}{\partial\theta}\left(\frac{1}{\cos\theta}\frac{\partial\psi}{\partial\theta}\right)-\frac{f(1-\kappa)}{a}\frac{\partial U}{\partial p}\frac{\partial \psi}{\partial\theta}+\frac{pf\cot\theta}{a^2}\frac{\partial U}{\partial p}\frac{\partial\psi}{\partial p}=\nonumber\\
% &+\frac{fp}{a\cos\theta}\frac{\partial^2\left(\overline{u'v'}\cos^2\theta\right)}{\partial\theta\partial p}+fp\cos\theta\frac{\partial^2\overline{u'\omega'}}{\partial p^2}-\frac{\cos\theta}{a^2}\frac{\partial}{\partial_\theta}\left(\frac{1}{\cos\theta}\frac{\partial\left(\overline{v'T'}\cos\theta\right)}{\partial\theta}\right) -\frac{\gamma R\cos\theta}{a}\frac{\partial^2\left(\overline{\omega'\theta'}\right)}{\partial\theta\partial p}\nonumber\\
%   &-fp\cos\theta\frac{\partial F}{\partial p}+\frac{R}{C_p}\frac{\cos\theta}{a}\frac{\partial Q}{\partial \theta}  \label{eq:hadley-drive}\\
%   &M_y=f-\frac{1}{a\cos\theta}\frac{\partial}{\partial\theta}(U\cos\theta)
% \end{eqnarray}
where $R$ is gas constant, $a$ is the earth radius, $\psi$ is the meridional streamfunction defined as below:
\begin{eqnarray}
  &\bar{v}=\frac{1}{\cos\theta}\frac{\partial\psi}{\partial p}\\
  &\bar{\omega}=-\frac{1}{a\cos\theta}\frac{\partial \psi}{\partial \theta}.
\end{eqnarray}

We solve for $\psi$ forced by each individual term on the right hand side (RHS) to decompose the meridional circulation. Since this is a Poisson equation, any initial $\psi$ will finally converge to the solution if it is updated using the tendency RHS$-$LHS.

 Note again that by invoking the thermal wind balance, we naturally have the time drift
terms in Eq.~\ref{eq:U-equation} and Eq.~\ref{eq:T-equation} canceled out without assuming $u(t+\delta t) - u(t) = T(t+\delta t) - T(t) =0$. This allows us to decompose the meridional circulation for one particular season. Also, because of the invoking of the thermal wind balance, the nonlinear advection terms of $\overline{u}$ and $\overline{T}$ (Eq.~\ref{eq:advection}) is also canceled to a large extent, as will be verified in the results section. 

The total meridional circulation is also diagnosed here from a different perspective, by taking vertical integral of meridional velocity in pressure coordinate.

\begin{eqnarray}
  \label{eq:total-psi}
  \psi=\frac{2\pi a\cos\theta}{g}\int^p_0~\bar{v}~dp'~ (\mathrm{unit: Sv}=10^9~\mathrm{kg/s}).
\end{eqnarray}
To match the units of the previous calculation, the solution of Eq.~\ref{eq:hadley-drive} is multiplied by a factor, $2\pi a/g$. %That the total of the forced streamfunctions equal the total streamfunction from Eq.~\ref{eq:total-psi} indicates a closed decomposition.
We then define the projected circulation index (PCI) to evaluate how much of the total circulation is driven by a specific forcing component, by projecting the solution of Eq.~\ref{eq:hadley-drive}, $\psi^{(i)}$ onto the normalized total streamfunction $\psi/|\psi|$ from Eq.~\ref{eq:total-psi}:
\begin{eqnarray}
  \label{eq:projected-circulation-index}
\Psi_{\mathrm{proj}}^{(i)}= \frac{\int\int \psi^{(i)}(\theta, p)\psi(\theta,p)~d\theta dp}{\sqrt{\int\int \psi^2(\theta,p)~d\theta dp}},
\end{eqnarray}
where $i$ denotes the component being evaluated. In the diabatic/adiabatic decomposition, $i$ could be the ``diabatic'' component forced by term associated with $Q$, or the ``adiabatic'' component forced by eddy transport terms and friction $F$  (see Eq.~\ref{eq:hadley-drive}), or ``advective'' component forced by the terms in Eq.~(\ref{eq:advection}). In the thermal/momentum decomposition, $i$ could be the ``thermal'' component forced by the terms originated from the temperature equation (including the ones associated with $\overline{v'T'},\ \overline{\omega'\theta'}$ and $Q$), or the ``momentum'' component forced by the terms originated from the momentum equation (including the ones associated with $\overline{u'v'}$, $\overline{u'\omega'}$ and $F$), or the ``advective'' component, whose definition is the same as before.

If removing the square root in the denominator of Eq.~(\ref{eq:projected-circulation-index}), each $\Psi_{\mathrm{proj}}^{(i)}$ would represent the percentage of total meridional streamfunction explained by a specific component, and the sum of them would be 1. We, instead, multiply an additional factor $\sqrt{\int\int \psi^2(\theta,p)~d\theta dp}$ to the above definition, so that PCIs have the units of a streamfunction. This way, the relative magnitude of PCIs associated with different physical processes for a given climate state represents the relative importance of these processes, and meanwhile, the PCIs for a specific physical process can be compared across different climate states to tell how the circulation strength changes.

Eq.~\ref{eq:hadley-drive} formulates a framework for evaluating the meridional circulation driven by any specific momentum or thermal forcing. Except the circulation driven by the advection term (Eq.~\ref{eq:advection}), solving it requires almost no knowledge of the climatological state $\bar{u}$ and $\bar{T}$. The physical picture is as follows. The planetary rotation creates a meridional gradient of angular momentum ($\bar{M}$), and latent heating release creates a vertical gradient of potential temperature ($\bar{\Theta}$). $\bar{M}$ and $\bar{\Theta}$ conservations constrain air parcels from moving in the meridional plane without external momentum or thermal forcing. The solution of Eq.~\ref{eq:hadley-drive} will be able to exactly counterbalance the external forcings, while satisfying both mass continuity and thermal wind balance. $\bar{u}$ and $\bar{T}$ are implicitly determined: one can substitute $\Psi$ back to Eq.~\ref{eq:T-equation} and Eq.~\ref{eq:U-equation}, and solve for $\bar{u}$ and $\bar{T}$, provided proper boundary conditions. For any given external forcing, there will be such a pair of $\bar{u}$ and $\bar{T}$, automatically satisfying the thermal wind relationship. When solving for a different external forcing, the corresponding $\bar{u}$ and $\bar{T}$ will also change. This is an indication that the decomposition is not ``clean''. Fully decomposing the interactions among the mean circulation, the eddies and the diabatic terms is not possible.

%The circulation forced by the advection term (Eq.~\ref{eq:advection}) can be interpreted as a self enhancement/suppression, as the circulation carries around air parcels with different angular momentum and potential temperature and breaks down the thermal wind balance.

%Alternatively, one could move the advection terms (Eq.~\ref{eq:advection}) to the left hand side and solve the elliptic equation. This is equivalent to define a new $M\times\Theta$ coordinate after considering the adjustment by the background wind $U$ and temperature $T$. However, the solution will contain the knowledge of background $U,\ T$; and the algorithm turns out to converge much slower with less accuracy, due to the scarceness of the $M$ contours. This is particular obvious for the solstice season circulation, as $M$ is almost conserved following the circulation. We therefore will stick to the previous method.

\section{Results}
\label{sec:results}

The results section is divided into two parts. We first show in section~\ref{sec:two-regime-behavior} that on high obliquity planets, the meridional circulation has two regimes: an adiabatic-dominant/momentum-driven regime in the limit of low insolation and fast rotation, and a diabatic-dominant/thermal-driven regime in the limit of high insolation and slow rotation. We then understand the circulation by comparing it against the circulation in the idealized dry dynamic core model study by \citet[][KCT]{Kang-Cai-Tziperman-2019:tropical} in section~\ref{sec:insights-from-dry}. Despite the sophisticated physics processes involved in the GCM here, the baroclinic eddies and the circulation driven in the adiabatic-dominant regime share many common characteristics with those in the dry model, which, in turn, was shown to be predicted by a 1D linear baroclinic instability model (KCT). We end with potentially observable diagnosis, including the upper atmospheric jet and the cloud distribution, are then discussed in section~\ref{sec:obs-evidence-regime-trans}. 

\subsection{Two regime behavior of the meridional circulation}
\label{sec:two-regime-behavior}

% motivation to separate adiabatic from diabatic
The adiabatic and diabatic processes respond to external forcings (such as orbital parameters) in different manners, despite the close coupling between them. This motivates us to distinguish between their roles in driving the meridional circulation. The diabatic processes include the radiative and latent heating, and the latter enhances with temperature and thus insolation. The adiabatic processes, on the other hand, include friction and eddy transports, whose pattern and amplitude can be strongly affected by the rotation rate. We here investigate how the adiabatically-driven and diabatically-driven meridional circulations change with insolation and rotation rate. 

%% Circulation at different insolations
\paragraph{The circulation at different insolations. }

 \begin{figure*}[!tbh]
 \centering
 \includegraphics[width=\textwidth]{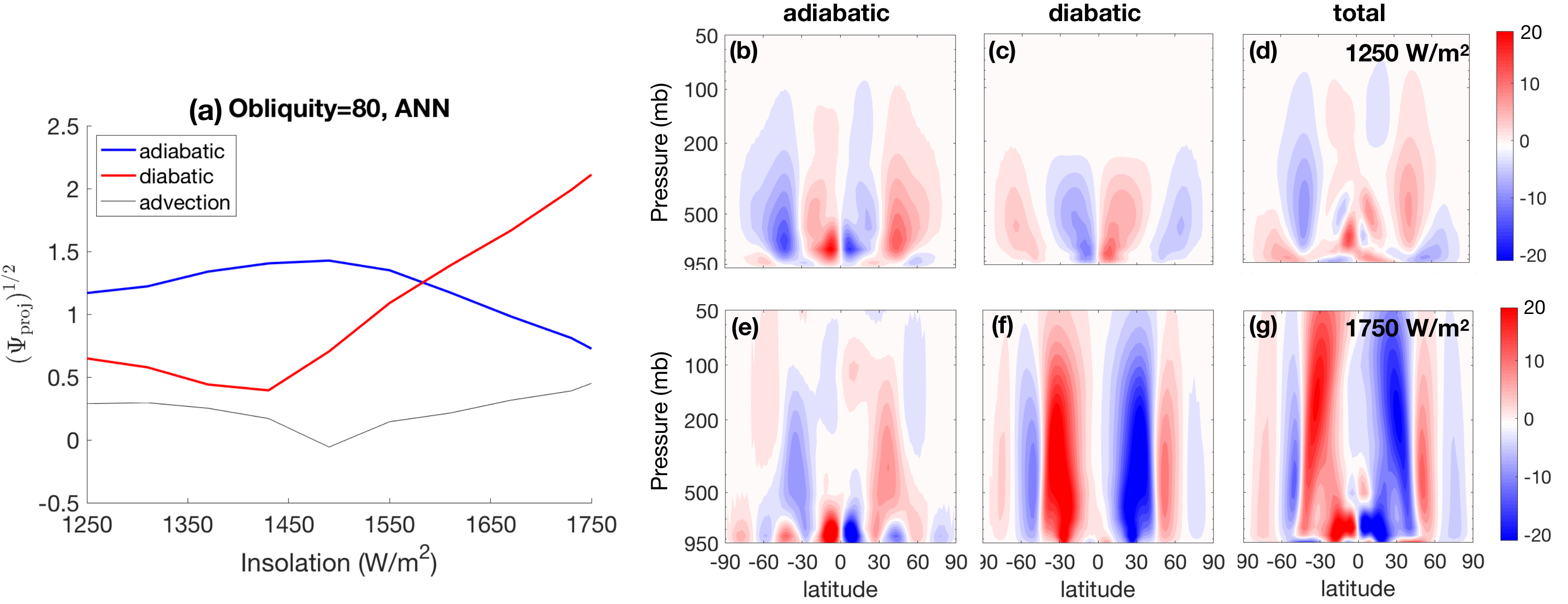}
 \caption{Regime transition of annual mean meridional circulation as varying insolation for OBL80-S experiment. (a) shows the projected circulation index (PCI, defined in Eq.~\ref{eq:projected-circulation-index}) as a function of insolation. The contribution from a specific component is measured by projecting the associated meridional streamfunction over the total streamfunction, and taking sum over all grid points (see methods section for details). We show the sign preserved square-root of PCIs ($\sqrt{|\mathrm{PCI}|}\cdot\mathrm{sign}[\mathrm{PCI}]$), in order to zoom in the regions with low PCI. Panel (b,e), panel (c,f), and panel (d,g) are the meridional circulation driven by adiabatic processes, diabatic processes and the total meridional streamfunction, plotted as a function of pressure and latitude. Panel (b,c,d) are for 1250 W/m$^2$ insolation, and panel (e,f,g) are for 1750 W/m$^2$. Red indicates clockwise circulation and blue indicates anticlockwise circulation.}
 \label{fig:regime-varyS0-obl80-ann}
\end{figure*}

\begin{figure*}[!tbh]
 \centering
 \includegraphics[width=\textwidth]{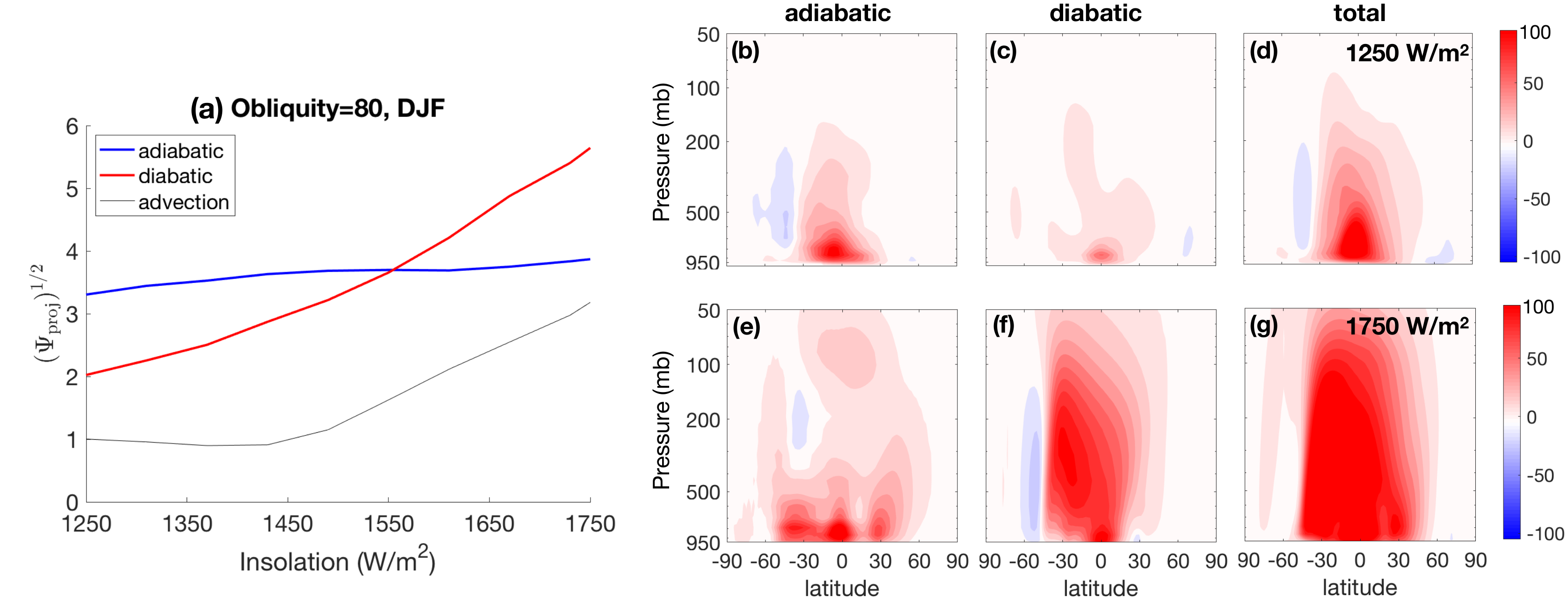}
 \caption{Same as Fig.~\ref{fig:regime-varyS0-obl80-ann}, but for the boreal winter season (DJF).}
 \label{fig:regime-varyS0-obl80-djf}
\end{figure*}

% two regime behavior for the high obliquity experiments with varying insolation
We measure the circulation contributed by diabatic and adiabatic processes with the projected circulation index (PCI, Eq.~\ref{eq:projected-circulation-index}) as insolation increases. The square root of PCIs of the diabatic and adiabatic circulations under high obliquity are shown in Fig.~\ref{fig:regime-varyS0-obl80-ann}a for the annual mean, and in Fig.~\ref{fig:regime-varyS0-obl80-djf}a for boreal winter (DJF) season. For both seasons, the meridional circulation has two regimes, with a transition near 1550 W/m$^2$. In the low insolation end, the PCI of the adiabatic component is around 4 times greater than that of the diabatic component, meaning that adiabatic processes contribute 4 times more circulation than diabatic processes, consistent with KCT. The dominance of the adiabatic-driven circulation can also be clearly seen in the circulation streamfunctions (panel b,c).
% As the insolation strenghthens, the diabatical PCI increases significantly, while the adiabatic PCI remains almost unchanged.
In the high insolation end, the ratio reverses -- diabatic processes contribute 4 times more than the adiabatic processes, as demonstrated in panel (e,f). %As the diabatic processes start to dominate, the tropical circulation transfers from one that is bottom-amplified and mostly thermally-indirect to one that is much more extended in the vertical direction and mostly thermally-direct. The circulation pattern will be discussed in more details in section~\ref{sec:insights-from-dry}.

\begin{figure*}[!tbh]
 \centering
 \includegraphics[width=\textwidth]{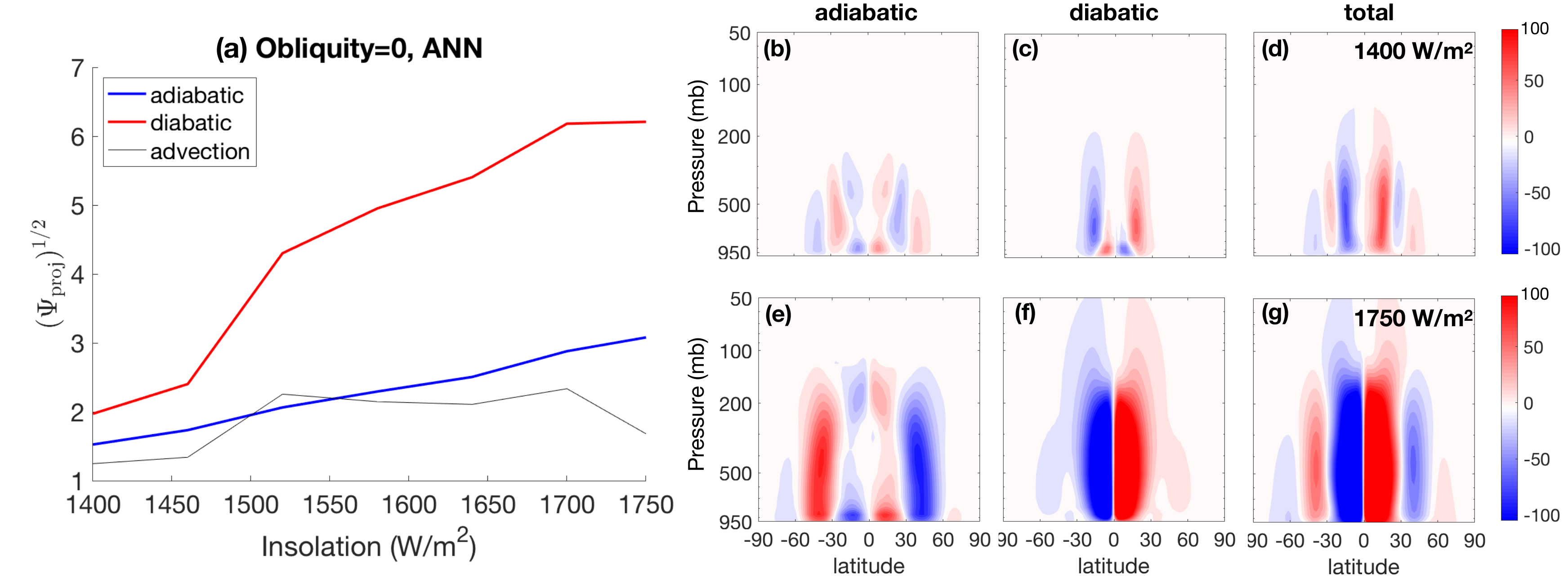}
 \caption{Same as Fig.~\ref{fig:regime-varyS0-obl80-ann}, but for the zero obliquity experiment (OBL0-S). Red indicates clockwise circulation and blue indicates anticlockwise circulation.}
 \label{fig:regime-varyS0-obl0-ann}
\end{figure*}

% for low obliquity 
Under low obliquity, the diabatically-driven circulation always dominates in the tropics and adiabatically-driven circulation dominant elsewhere (Fig.~\ref{fig:regime-varyS0-obl0-ann}a). This is consistent with the understanding of the atmospheric general circulation on Earth \citep{Pfeffer-1987:comparison}, but contradicts the results in dry dynamic core study (KCT), where the adiabatic processes were found to be dominant everywhere. This inconsistency occurs probably because the latent heating is fully represented in a Held-Suarez type restoring physics.
%Since this latent heating is concentrated near the equator, where the effect of rotation is weak (quantified by a large Rossby number), it drives a strong thermally-direct circulation there. Adiabatic-driven circulation dominates in the extratropics (Ferrel cell), while the diabatic-driven one dominates in the tropics (Hadley cell), consistent with the understanding of the atmospheric general circulation on Earth.

% diabatic component enhances with insolation, but the adiabatic remains constant
One common feature that is evident in all cases is that the diabatic-driven circulation enhances significantly with insolation, whereas the amplitude of the adiabatic-driven circulation remains almost constant. Adiabatic processes include surface friction and eddy heat/momentum transport, none of which is directly affected by the insolation, explaining the lack of sensitivity of adiabatic-driven circulation to insolation. However, diabatic processes, composed of radiative heating/cooling and latent heating, are highly sensitive to insolation. Radiative heating/cooling rises proportionally to the insolation magnitude, given a fixed latitudinal insolation profile. Latent heating surges up with temperature following the Clausius-Clapeyron relation.

%% Circulation at different rotation rates
\paragraph{The circulation at different rotation rates. }

\begin{figure*}[!tbh]
 \centering
 \includegraphics[width=\textwidth]{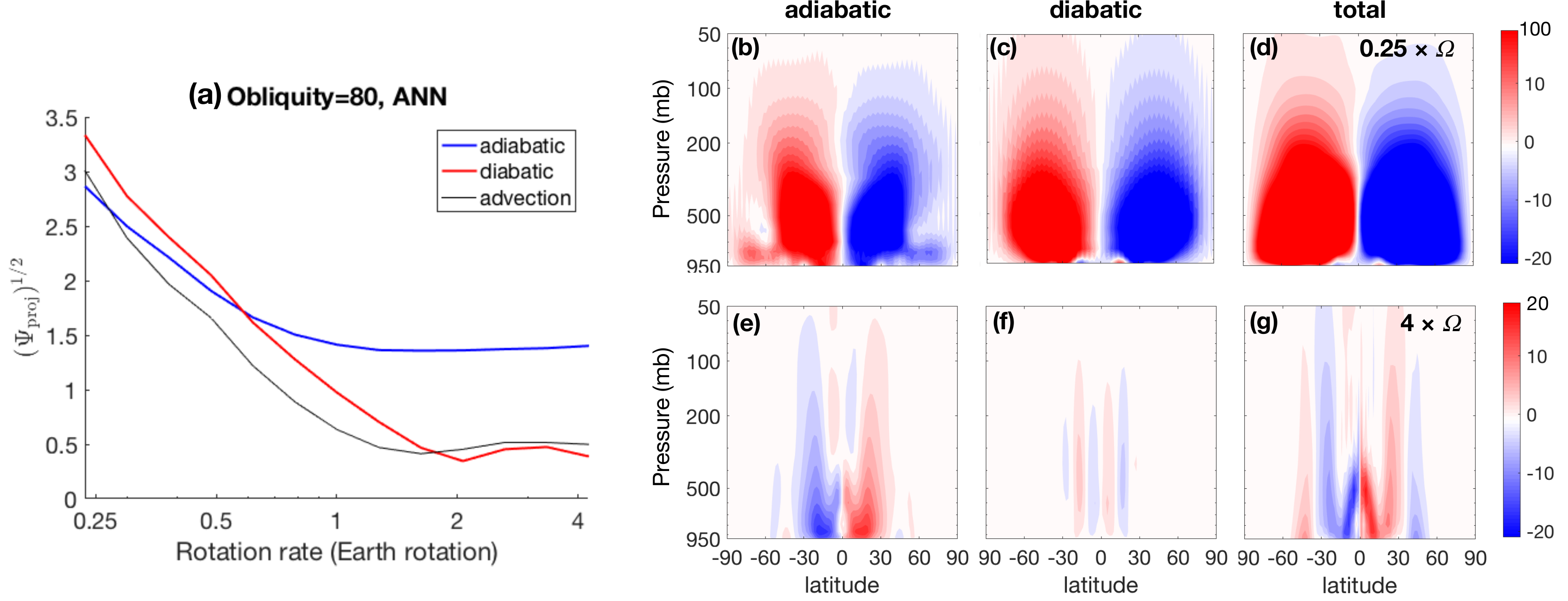}
 \caption{Same as Fig.~\ref{fig:regime-varyS0-obl80-ann}, except self rotation rate instead of insolation is varying here. Red indicates clockwise circulation and blue indicates anticlockwise circulation.}
 \label{fig:regime-varyOm-obl80-ann}
\end{figure*}

\begin{figure*}[!tbh]
 \centering
 \includegraphics[width=\textwidth]{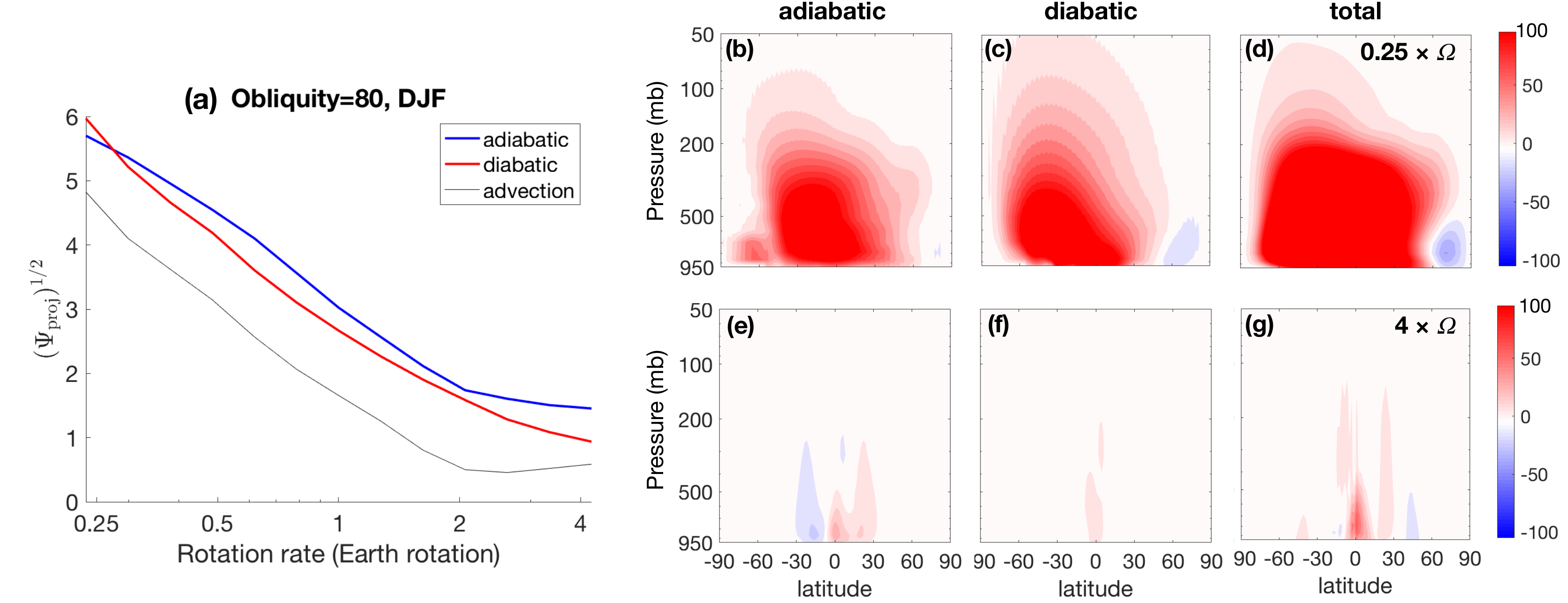}
 \caption{Same as Fig.~\ref{fig:regime-varyS0-obl80-djf}, except self rotation rate instead of insolation is varying here. Red indicates clockwise circulation and blue indicates anticlockwise circulation.}
 \label{fig:regime-varyOm-obl80-djf}
\end{figure*}

\begin{figure*}[!tbh]
 \centering
 \includegraphics[width=\textwidth]{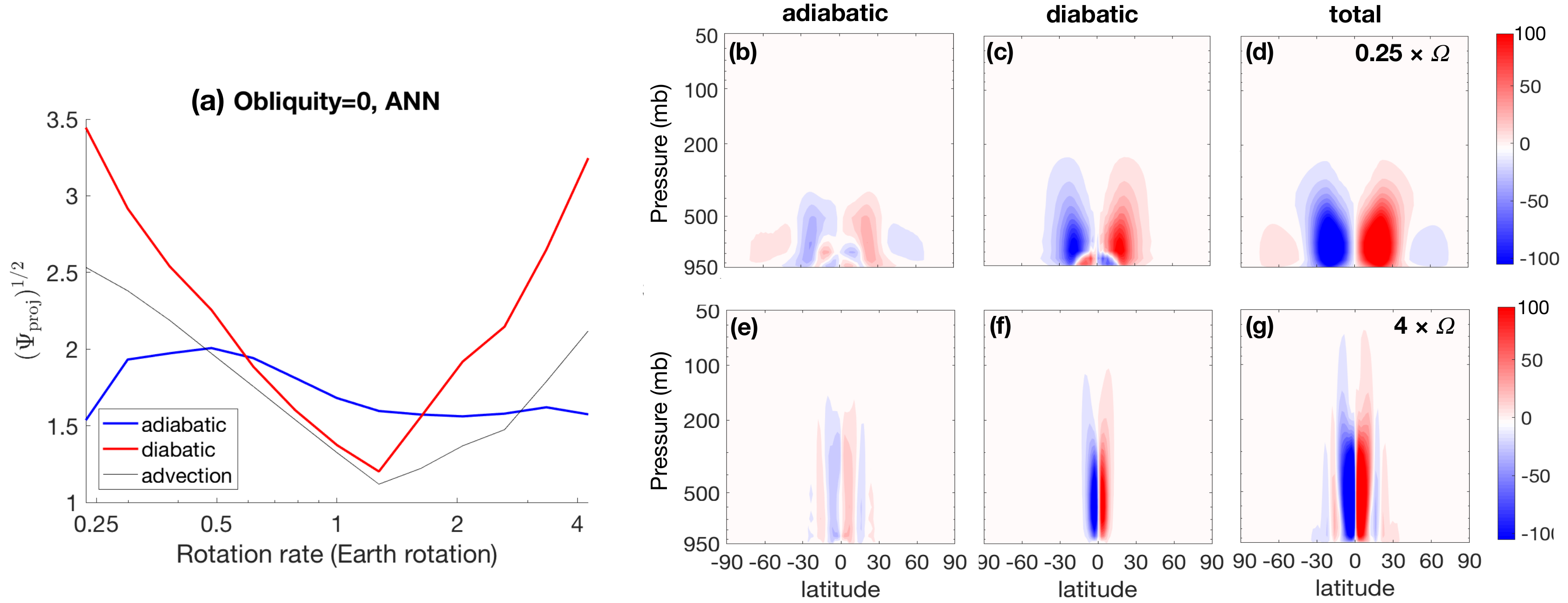}
 \caption{Same as Fig.~\ref{fig:regime-varyS0-obl0-ann}, except self rotation rate instead of insolation is varying here. Red indicates clockwise circulation and blue indicates anticlockwise circulation.}
 \label{fig:regime-varyOm-obl0-ann}
\end{figure*}

We vary the planet's rotation rate, with insolation fixed at 1360 (1550) W/m$^2$ for the low (high) obliquity scenario, as this is the levels where the adiabatic and diabatic driven circulations have similar magnitudes in OBL0-S and OBL80-S. The PCIs for adiabatic and diabatic driven circulation are shown in Fig.~\ref{fig:regime-varyOm-obl80-ann}a and Fig.~\ref{fig:regime-varyOm-obl80-djf}a for the high obliquity annual-mean and DJF season, and in Fig.~\ref{fig:regime-varyOm-obl0-ann} for the low-obliquity circulation.

% common features: circulation weaken with rotation rate. 
The meridional circulation should get weaker and narrower with rotation rate, following Held-Hou scaling \citep{Held-Hou-1980} or Held-2000 scaling \citep{Held-2000:general}. %With faster rotation rate, air parcels gain westerly speed more easily as they move meridionally conserving angular momentum, and thus, thermal wind balance with a specific meridional temperature gradient can be achieved with less meridional displacement.
% high obliquity, no clear regime transition when changing rotation
% thermally indirect => thermally direct as rotation decreases
As expected, under high obliquity, both the diabatically and adiabatically driven circulations weaken with rotation rate (Fig.~\ref{fig:regime-varyOm-obl80-ann}a and Fig.~\ref{fig:regime-varyOm-obl80-djf}a). Thus the regime transition is less clear compared to that in OBL80-S. Transition disappears for the DJF meridional circulation (Fig.~\ref{fig:regime-varyOm-obl80-ann}). %Together with the regime transition is a transition from a thermally-direct circulation pattern to a thermally indirect one. More discussion of the high obliquity circulation pattern is in section~\ref{sec:insights-from-dry}.

% low obliquity
Under low obliquity, the circulation response to rotation rate change is not monotonic. The diabatic-driven circulation, in particular, first weakens then strengthens, with a minimum around 1 times earth rotation rate (Fig.~\ref{fig:regime-varyOm-obl0-ann}a). The adiabatic-driven circulation responds in a much subtler manner. The weakening of circulation with rotation rate is expected, as explained above. The strengthening instead is associated with the low-latitude melting. Fast rotation prohibits poleward heat transport beyond tropics (not shown), and leads to surface warming in the low latitudes, which in turn melts the tropical sea ice, triggers convections and strengthens the circulation within tropics.

% advective term, budget closed, momentum/thermal decomposition.
The PCI of advection terms (Eq.~\ref{eq:advection}) is plotted as a thin black curve in panel (a) of Figs.~\ref{fig:regime-varyS0-obl80-ann}-\ref{fig:regime-varyS0-obl0-ann}. It is substantially smaller than either diabatic or adiabatic PCI for most circumstances, and it is therefore negligible. We have examined that the sum of the circulation forced by each individual component reproduce the total meridional circulation diagnosed by vertically integrating $\overline{v}$ (see the right two columns of Fig.~S7-12 in the supplementary figure file which is available from \url{https://www.dropbox.com/s/nsext5lc90hw6qe/supplementary_figures_high_obliquity_S0_Omega3.pdf?dl=0} or from Harvard dataverse \url{https://doi.org/10.7910/DVN/3JVBOS}).

Another interesting decomposition is to separate the circulation driven by eddy momentum transports and friction from that driven by thermal effects (eddy heat transports and diabatic heating). The same figures for the momentum/thermal decomposition is available \url{https://www.dropbox.com/s/nsext5lc90hw6qe/supplementary_figures_high_obliquity_S0_Omega3.pdf?dl=0} (Fig.~S1-6), all of the above conclusions still hold, except that the momentum driven circulation tends to play a more important role in the annual mean than during DJF.
%This may be because the thermal component driven by the asymmetric heating in the two hemispheres could be largely canceled in the annual mean, whereas the eddy transports and the surface friction pattern remain similar all year round (see the discussion of Fig.~\ref{fig:eddy-transport-obl80-dryrev} in section~\ref{sec:insights-from-dry}).

\subsection{Understanding the circulation structure changes near the transition point}
\label{sec:insights-from-dry}

The circulation regime transition under high obliquity is accompanied by the changes of the circulation structure. In this subsection, we try to understand the dynamics driving the different circulations. We particularly focus on the adiabatic/momentum driven component, because it shows richer behavior than simply rising at warm latitudes and sinking at cold latitudes, as it is for the diabatic/thermal component. The main message here is that the eddies and the meridional circulation in the high obliquity scenario resemble those in a dry dynamic core model forced by a reversed meridional temperature gradient used in KCT, and thus the dry model provides useful insights to understand the high obliquity climate. 

%They investigated the basics of high-obliquity circulation by simply reversing the meridional temperature gradient, motivated by a coupled ocean-atmosphere model simulation result that the poles are warmer than the equator all year round \citep{Ferreira-Marshall-OGorman-et-al-2014:climate}. 

\paragraph{Circulation structure's response to insolation. }

\begin{figure*}[!tbh]
 \centering
 \includegraphics[width=\textwidth]{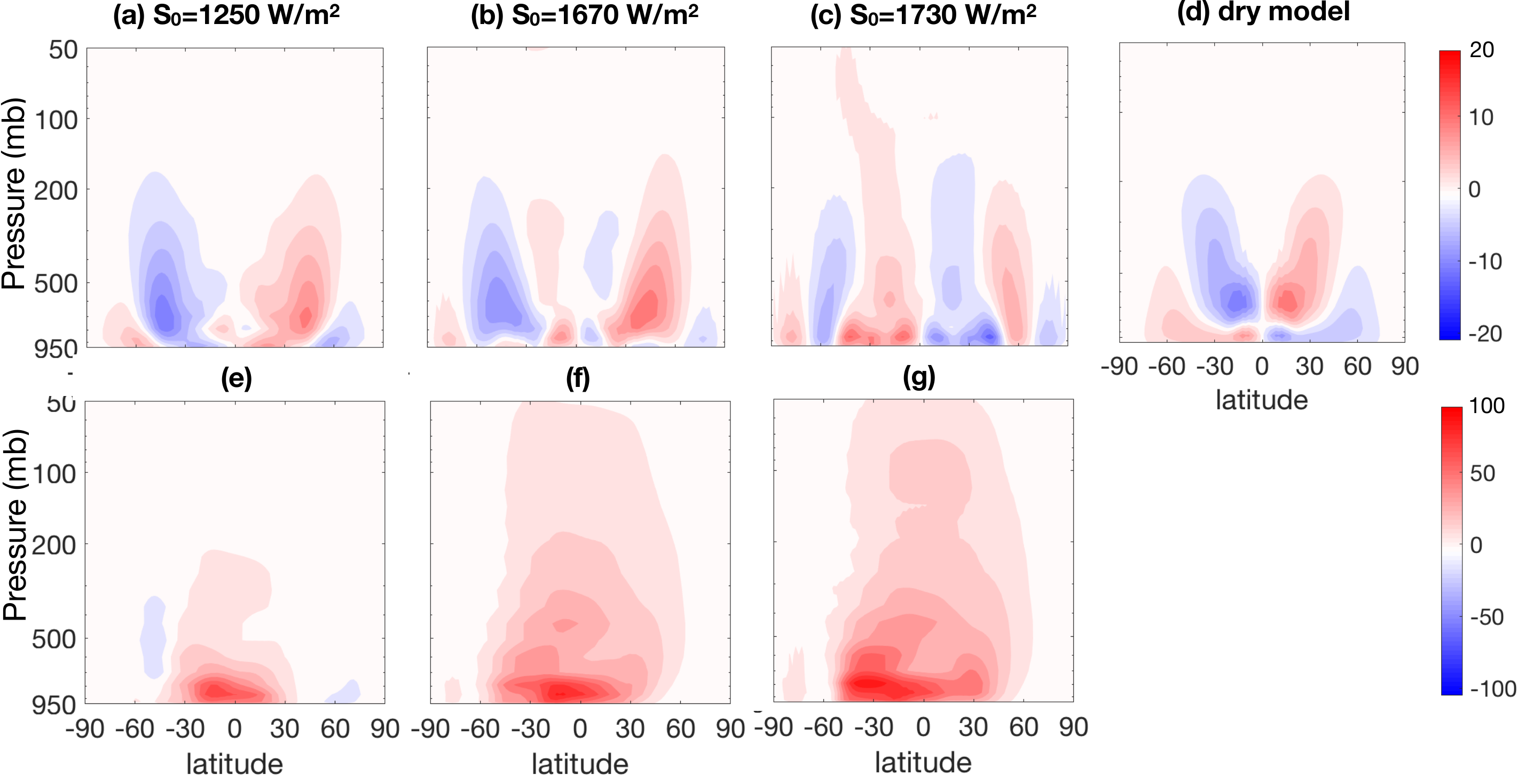}
 \caption{Momentum driven meridional streamfunction under different insolations. From (a) to (c), shown are for $S_0=1250,\ 1670,\ 1730 \ W/m^2$ in OBL80-S, and panel (d) is the total meridional streamfunction in the dry dynamic core model with a reversed meridional temperature gradient. (e-g) are the same as (a-c) except for DJF season. Momentum driven circulation represents the circulation driven by meridional eddy momentum transport $\overline{u'v'}$, vertical eddy momentum transport $\overline{u'\omega'}$, and friction.}
 \label{fig:momentum-psi-diff-S0}
\end{figure*}

% OBL=80 annual mean: linkage to the reverse gradient paper
Fig.~\ref{fig:momentum-psi-diff-S0}(a-c) show the momentum-driven\footnote{There is no unique ``correct'' way to decompose the circulation. To avoid the cancellation between the eddy heat transport and the diabatic processes, the meridional circulation here is decomposed into the momentum-driven and thermally-driven components to avoid the cancellation between the eddy heat transport and the diabatic processes.} circulation under different insolations. In panel (d), the total meridional circulation in the reverse gradient dry model used in KCT is reprinted to be compared with the circulation the realistic high-obliquity simulations here. Below 1670 W/m$^2$, the momentum-driven circulation under high obliquity remains remarkably similar to that in the dry model. Both are characterized by a thermally-indirect cell within 45N/S, and a thermally-direct cell in high latitudes constrained near the surface. Since the momentum-driven circulation explains most parts of the total meridional circulation below 1550 W/m$^2$ insolation, the total meridional circulations in this regime also resemble that in the dry model. The thermally-indirect cell does not exist in most of the high obliquity simulations in \citet{Linsenmeier-Pascale-Lucarini-2015:climate}, probably because they constrain their simulation near the outer edge of the habitable zone, or because the baroclinic eddy activity is underestimated due to the low resolution of their model. 
decomposition.

\begin{figure*}[!tbh]
 \centering
 \includegraphics[width=0.8\textwidth]{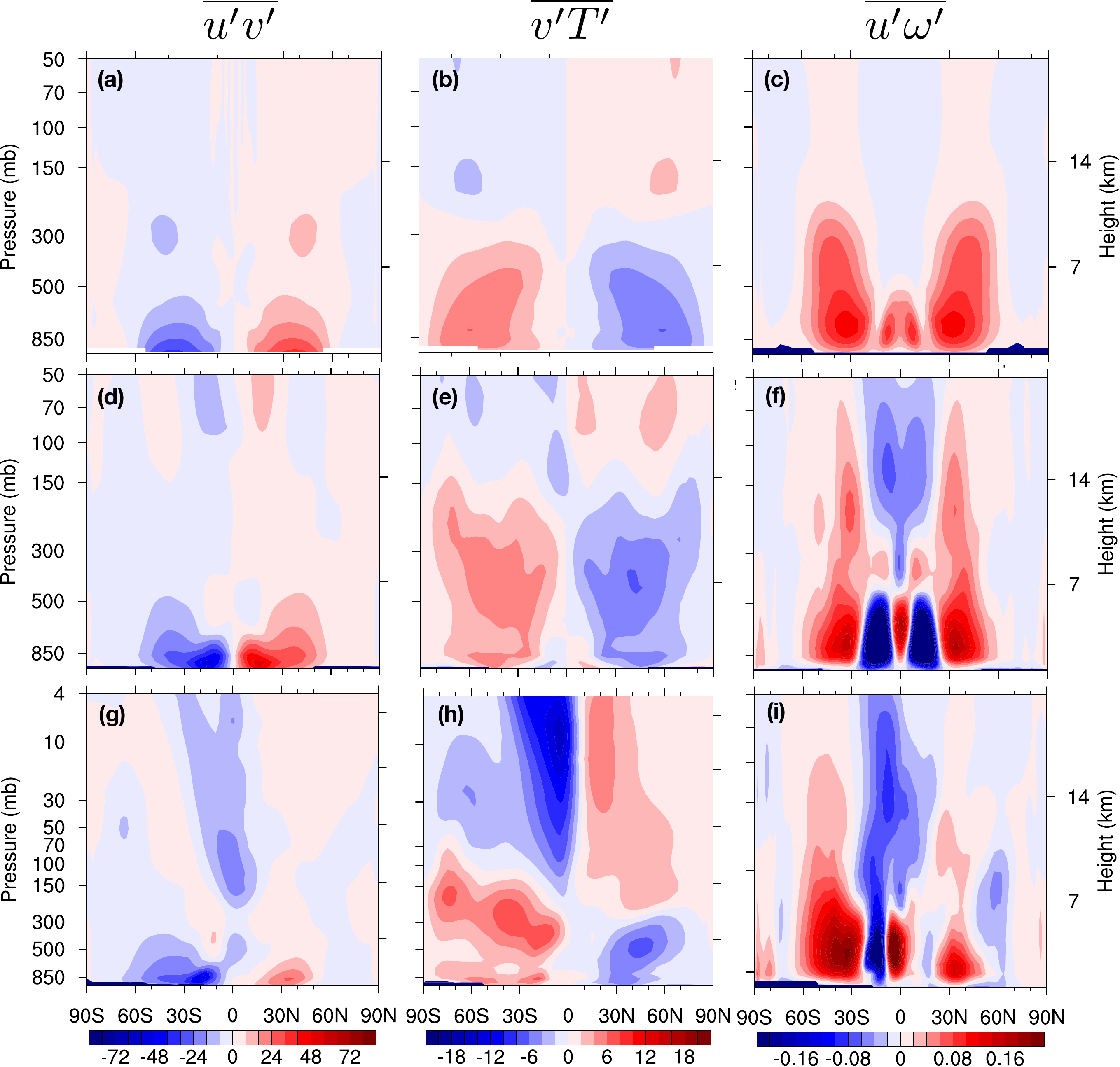}
 \caption{Eddy heat/momentum transports for (a,b,c) the dry dynamic model used in KCT, forced with reversed meridional temperature gradient, (d,e,f) the annual mean of 80 degree obliquity experiment, and (g,h,i) the DJF climatology of 80 degree obliquity experiment. All experiments are forced by $S_0=1360~W/m^2$. From left to right, shown are meridional eddy momentum transport $\overline{u'v'}$, meridional eddy heat transport $\overline{v'T'}$, and vertical eddy momentum transport $\overline{u'\omega'}$. }
 \label{fig:eddy-transport-obl80-dryrev}
\end{figure*}

% because eddy features are similar
The similarity in the meridional circulation stems from the similarity of baroclinic eddies, because eddies dominantly drive the circulation\footnote{Surface friction can also drive meridional circulation, however, it is slaved by the eddies. The surface wind will keep evolving until the surface friction can exactly counterbalance the column integrated eddy momentum convergence $\int\partial_y\overline{u'v'}~dp$.}

Shown in Fig.~\ref{fig:eddy-transport-obl80-dryrev}(d-f) and (g-i), respectively, are the annual mean and DJF climatologies of $\overline{u'v'}$, $\overline{v'T'}$ and $\overline{u'\omega'}$ under high obliquity forced by $1360~W/m^2$ insolation. They share almost all of the eddy characteristics with the dry model (Fig.~\ref{fig:eddy-transport-obl80-dryrev}a-c). The similarity indicates that eddy characteristics are not qualitatively affected by the newly included physics processes (radiation, hydrological cycle etc.) and the seasonal cycle in ExoCAM. For both ExoCAM and the dry model: $\overline{u'v'}$ is concentrated near the surface rather than the upper troposphere as in the zero-obliquity situation (Fig.~\ref{fig:eddy-transport-obl0}a). $\overline{v'T'}$ switches sign at the mid-troposphere, different from the zero-obliquity experiment which has two peaks of the same sign (Fig.~\ref{fig:eddy-transport-obl0}b). $\overline{u'\omega'}$ is mostly positive rather than the opposite as in the zero-obliquity situation.

KCT attempts to understand the eddy characteristics using a 1D linear baroclinic instability model, and explain why the meridional circulation is thermally-indirect, shallow and weak with a reversed meridional temperature gradient using these eddy characteristics. As discussed in KCT, eddies converge westerly momentum to the mid-latitude baroclinic zone, forming surface westerlies there and surface easterlies elsewhere. The Ekman friction acting on the tropical surface easterly leads to an upward motion and a thermally-indirect circulation in the tropics. Acceleration due to eddy momentum transport always counterbalances the surface friction; they together bound the vertical extension of the meridional circulation. Since the most unstable baroclinic mode is bottom-amplified in the climatology of the reversed gradient case, the circulation is shallower compared to the case with a normal meridional temperature gradient. Readers are referred to KCT for further details. 

% eddy behavior beyond 1670
The baroclinic eddy characteristics remain similar to that in the dry model until 1670 W/m$^2$, far beyond the circulation transition point 1550 W/m$^2$; so do the circulation driven by these eddies. Beyond 1670 W/m$^2$, the momentum driven circulation starts to become thermally direct from the bottom (Fig.~\ref{fig:momentum-psi-diff-S0}c). This circulation change coincides with the changes of the baroclinic eddy features. The whole pattern of $\overline{u'\omega'}$ shifts upward (not shown), consistent with \citet{Singh-OGorman-2012:upward}. As a result, the westerly momentum accumulation in the low latitudes peaks around 500 mb instead of close to the surface, pushing the 500 mb air parcels equatorward through Coriolis force, and leading to equatorial sink below that level.

%% OBL=80 DJF: vertical extension
The solstice meridional circulation under high obliquity is a cross-equator cell rising in the summer hemisphere and sinking in the winter hemisphere, regardless of the insolation. Although the circulation direction does not change, its vertical extension increases with insolation as the diabatic/thermal driven circulation expands to higher altitudes and starts to dominate (Fig.~\ref{fig:regime-varyS0-obl80-djf}c,f). The adiabatic/momentum driven circulation, on the other hand, is always limited below 300 mb (see Fig.~\ref{fig:regime-varyS0-obl80-djf}b,e for the adiabatic component and Fig.~\ref{fig:observable-obl80-S}e-g for momentum driven component), until 1670 W/m$^2$. Therefore, the vertical extension of the meridional circulation suddenly starts to rise at 1550 W/m$^2$, the transition point, where the diabatic/thermal component begins to dominate.

% DJF eddy features
The shallowness of the adiabatic/momentum driven circulation has its root in the shallowness of the baroclinic eddy structure, which can be seen from the meridional eddy momentum transport $\overline{u'v'}$ in Fig.~\ref{fig:eddy-transport-obl80-dryrev}g. The DJF eddy heat transport $\overline{v'T'}$ and vertical eddy momentum transport $\overline{u'\omega'}$ are shown together in panel (h,i). Even though the insolation distribution during DJF is drastically different from that in annual-mean, the eddies characteristics are not qualitatively different from the annual mean (panel d-f), except that the summer hemisphere has a stronger eddy activity. This is not surprising given that the meridional surface temperature gradient remains reversed in all seasons \citep{Ferreira-Marshall-OGorman-et-al-2014:climate}, and that the summer hemisphere has a stronger meridional insolation gradient and hence a stronger baroclinicity.

%% OBL=0 ANN
%As for zero obliquity experiment forced by increasing insolation (OBL0-S), the total meridional circulation gets wider, taller and stronger, while remaining the three-cell structure (Fig.~\ref{fig:regime-varyS0-obl0-ann}d,g). The tropical Hadley cell enhancement is contributed by the diabatic/thermal component, while the mid-latitude Ferrel cell enhancement is by the adiabatic/momentum component. These changes have also been found the projection for a warmer earth \citep{Seidel-Fu-Randel-et-al-2008:widening, Lorenz-DeWeaver-2007:tropopause, Hartmann-Larson-2002:important, Thompson-Bony-Li-2017:thermodynamic, Hu-Huang-Zhou-2018:widening}.
%In a global warming scenario, the Hadley cell and Ferrel cell are expected to expand vertically \citep{Thompson-Bony-Li-2017:thermodynamic, Lorenz-DeWeaver-2007:tropopause}, and meridionally \citep{Seidel-Fu-Randel-et-al-2008:widening, Lorenz-DeWeaver-2007:tropopause, Hu-Huang-Zhou-2018:widening}, if assuming that the anvil temperature is fixed \citep{Hartmann-Larson-2002:important}, and that the Hadley's width and depth change proportionally \citep{Held-2000:general}.

\paragraph{Circulation structure's response to rotation rate. }

\begin{figure*}[!tbh]
 \centering
 \includegraphics[width=0.8\textwidth]{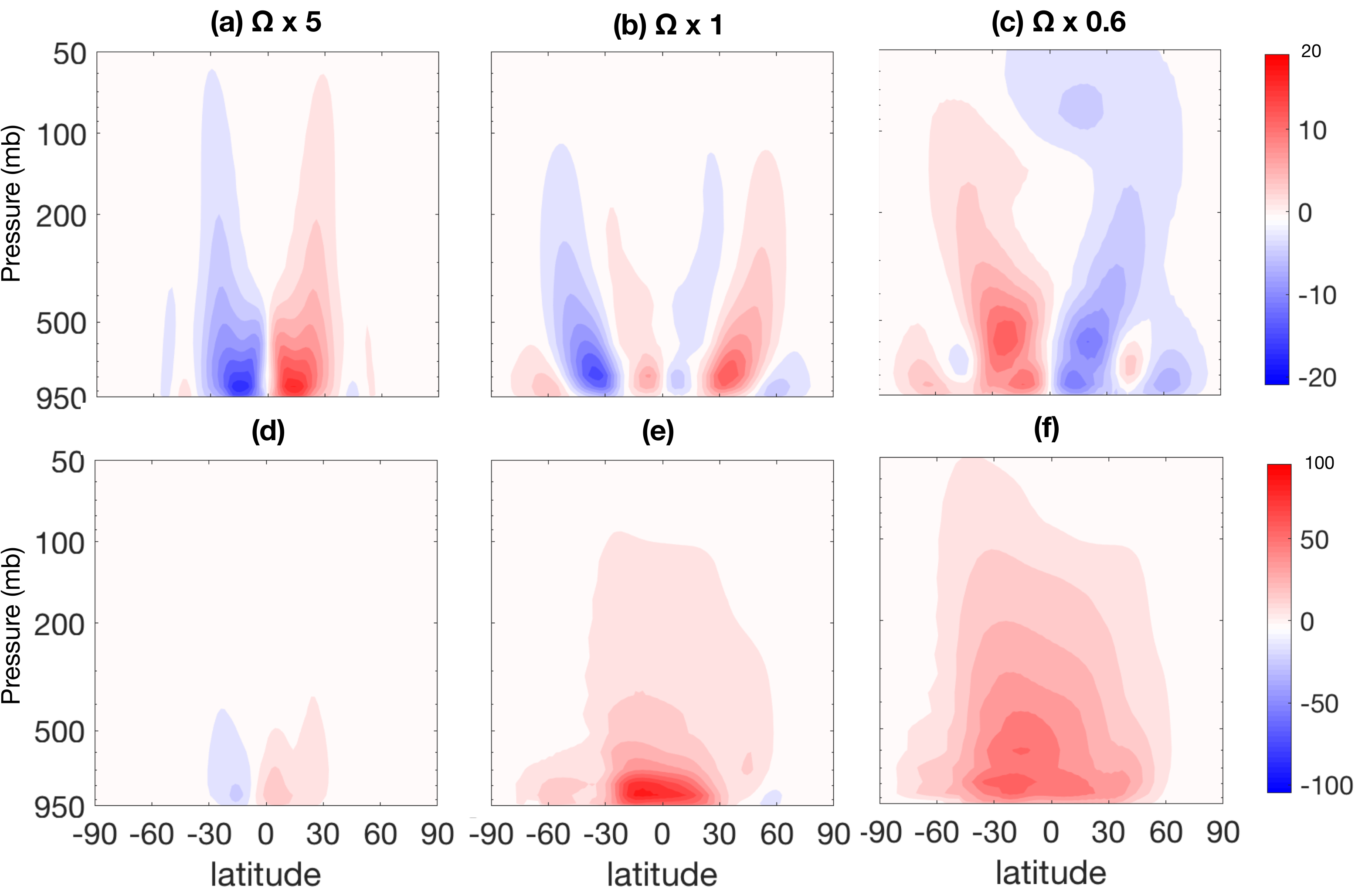}
 \caption{Same as Fig.~\ref{fig:momentum-psi-diff-S0}, but for different rotation rates. From left to right shown are $\Omega=5,\ 1,\ 0.6$ /earth day in OBL80-$\Omega$. Upper panels are for annual mean, and lower panels are for DJF season.}
 \label{fig:momentum-psi-diff-Om}
\end{figure*}

The meridional circulation under high obliquity narrows with the rotation rate, as seen in many low obliquity studies \citep[e.g.,][]{Kaspi-Showman-2015:atmospheric}. Concurrently, the annual-mean meridional circulation also becomes more thermally-direct (Fig.~\ref{fig:regime-varyOm-obl80-ann}). This is partially due to the rapid strengthening of the diabatic/thermal circulation component, and partially due to the transition of the adiabatic/momentum driven circulation from a thermally-indirect one into a thermally direct one, which is shown in Fig.~\ref{fig:momentum-psi-diff-Om}(a-c). At 5 times Earth rotation rate, the circulation is completely thermally-indirect (i.e., rising at the cold, sinking at the warm). As rotation rate decreases, circulation expands meridionally, giving space to a thermally-direct circulation in the tropics. When the rotation rate drops below 0.6 times earth value, the circulation becomes almost completely thermally-direct. This change is mostly due to the reversal of vertical momentum transport $\overline{u'\omega'}$ (not shown). At slower rotation rate, the upward westerly momentum transport in the tropics (Fig.~\ref{fig:eddy-transport-obl80-dryrev}f) strengthens and expands, driving a circulation that sinks at the equator.

\subsection{Observable evidence for the regime transition under high obliquity}
\label{sec:obs-evidence-regime-trans}

The regime transition seen in the high obliquity simulations (OBL80-S and OBL80-$\Omega$) has observable consequences.

\paragraph{Regime transition due to insolation. }

\begin{figure*}[!tbh]
 \centering
 \includegraphics[width=\textwidth]{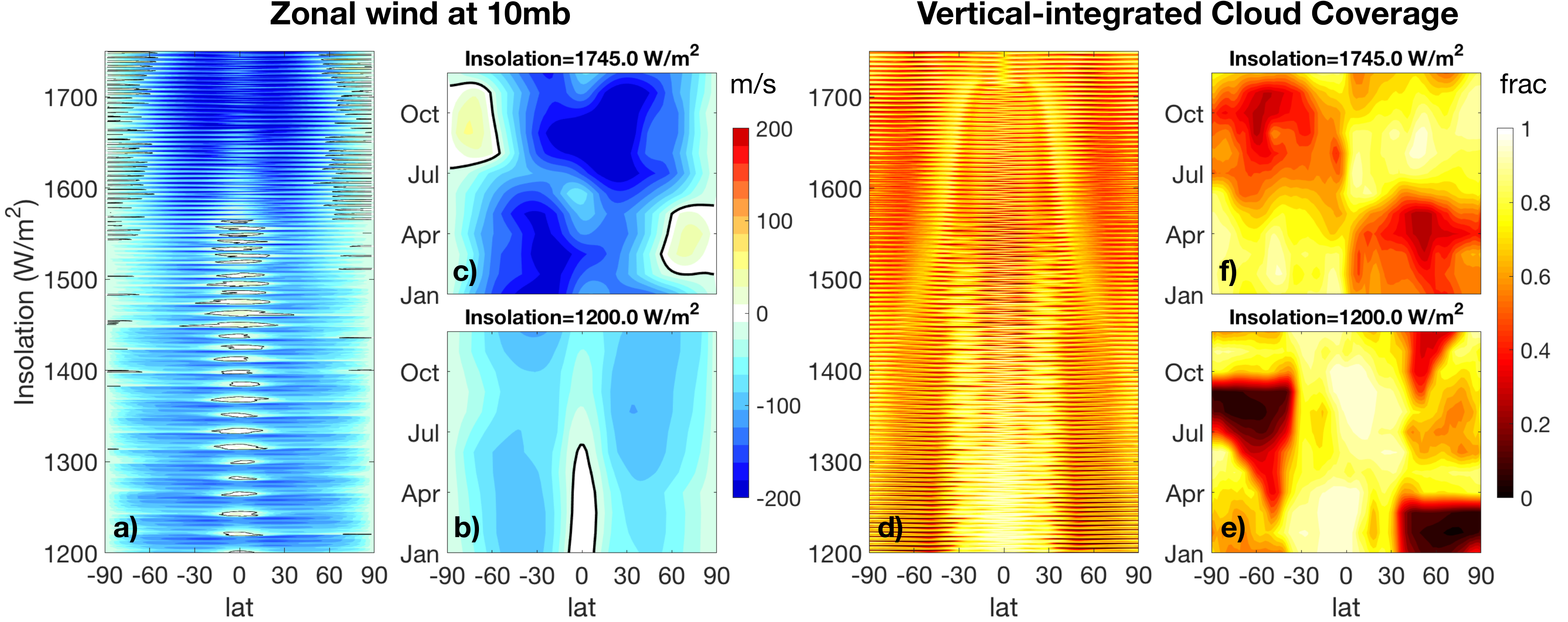}
 \caption{The observables for the regime transition in OBL80-S. (a-c) are for the 10 mb zonal wind, and (d-f) are for the vertically-integrated cloud cover. (a) and (d) show the whole progression of the zonal wind and cloud as insolation increases. While, (b,e) and (c,f) zoom in and show the seasonal cycle at the beginning and the end of the OBL80-S simulation, which corresponds to 1250 W/m$^2$ and 1750 W/m$^2$ insolation. Zero wind speed is denoted with black contours in (a-c).}
 \label{fig:observable-obl80-S}
\end{figure*}

% 10 mb zonal wind for OBL80-S
Fig.~\ref{fig:observable-obl80-S}a shows the 10 mb zonal wind as a function of latitude and insolation. Zero wind speed is denoted by a black contour. Below the transition point of 1550 W/m$^2$, the extratropical zonal wind is mostly easterly, while the equatorial zonal wind in the upper atmosphere switches direction with season: westerlies prevail the first half of a year and easterlies prevail the second half (Fig.~\ref{fig:observable-obl80-S}b). Beyond the transition point, upper air easterlies get stronger in most latitudes except the polar regions (Fig.~\ref{fig:observable-obl80-S}c). The strengthening of the tropical easterly across the transition point is expected given a more thermally-direct meridional circulation. As the poleward flow in the tropical upper atmosphere is replaced by an equatorward flow (Fig.~\ref{fig:regime-varyS0-obl80-ann}g), extratropical air parcels with low angular momentum is pumped toward the equator, accelerating the equatorial air near the tropopause toward the west. Starting from a more easterly base, the wind above will also be more easterly with a given meridional temperature gradient (the thermal wind balance constrains the vertical wind shear), unless there is a compensating momentum flux converging westerly momentum back to the equator again. As demonstrated by the analysis in \citet{Showman-Fortney-Lewis-et-al-2012:doppler}, different upper air wind patterns may be distinguishable in the transit phase curve, especially with the help of the future missions, such as James Webb Space Telescope, which can better resolve the infrared spectrum.

% cloud cover for OBL80-S
The vertically integrated cloud cover is shown in Fig.~\ref{fig:observable-obl80-S}d. Below the transition point, the cloud cover is high (greater than 80\%) all year round (Fig.~\ref{fig:observable-obl80-S}e). The highest cloud cover is located in the subtropics, supported by the upward motion there (Fig.~\ref{fig:regime-varyS0-obl80-ann}g). Only outside the tropics, does cloud cover show a strong seasonal variation, oscillating between almost 0\% and 80\%. Beyond the transition point, all latitudes show a strong seasonal variation, including the tropics  (Fig.~\ref{fig:observable-obl80-S}f). In other words, the tropical cloud cover drops. This is expected given that the equatorial vertical motion turns from upward to downward (Fig.~\ref{fig:regime-varyS0-obl80-ann}g). The peaks of the cloud cover through all seasons now move to 45N/S, happening after the summer solstice. This temporal high cloud cover near 45N/S is supported by the strong upward motion there (Fig.~\ref{fig:regime-varyS0-obl80-djf}g). Cloud cover may be detected through the albedo measurement \citep[e.g.,][]{Demory-Wit-Lewis-et-al-2013:inference}.

\paragraph{Regime transition due to rotation rate. }

\begin{figure*}[!tbh]
 \centering
 \includegraphics[width=\textwidth]{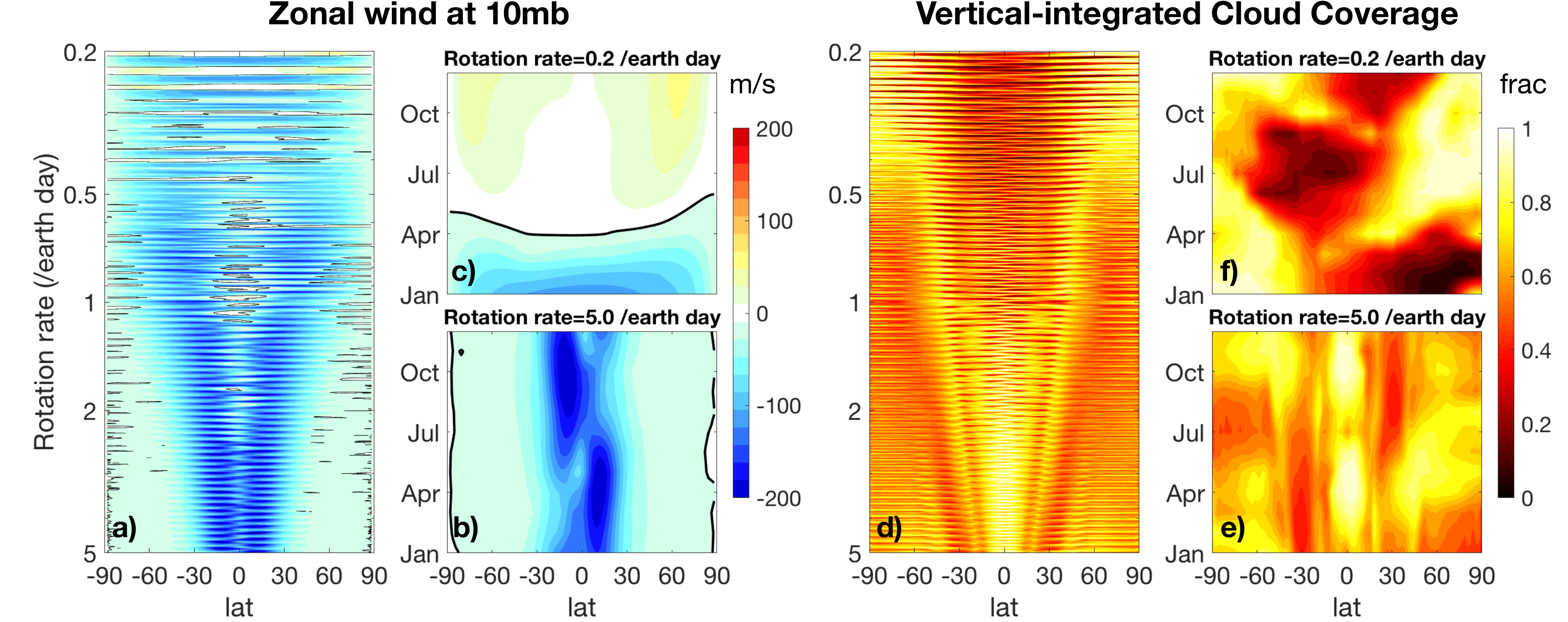}
 \caption{Same as Fig.~\ref{fig:observable-obl80-S}, except for OBL80-$\Omega$.}
 \label{fig:observable-obl80-Om}
\end{figure*}

% 10 mb zonal wind for OBL80-Om
Similar regime transition of the annual-mean meridional circulation is also seen when changing rotation rate. Fig.~\ref{fig:observable-obl80-Om}a shows the 10 mb zonal wind as a function of rotation rate and latitude. As the meridional circulation expands meridionally (Fig.~\ref{fig:regime-varyOm-obl80-ann}), the wind pattern also expands. In addition, the tropical easterly weakens, and finally starts to oscillate between easterly and westerly when the rotation rate is lower than that of the earth (Fig.~\ref{fig:observable-obl80-Om}a upper parts). These changes can be understood as follows. As circulation becomes more thermally-direct at a slower rotation rate, more extratopical air with low angular momentum is transported equatorward. However, this does not lead to a stronger equatorial easterly as it does in OBL80-S, because the meridional gradient of the planetary angular momentum, which is induced by the self rotation, also decreases proportionally. As a result, the wider, stronger and thermally-direct meridional circulation at slow rotation rate ends up transporting less easterly momentum equatorward. This may allow distinguish whether a thermal indirect (direct) circulation induced by low (high) insolation or by fast (slow) rotation rate.  Seasonal cycle of the upper atmospheric zonal wind is shown in Fig.~\ref{fig:observable-obl80-Om}(b,c) for the two most extreme conditions. In the fast rotating end, easterly wind is constrained within 30N/S, and the speed peaks in the late spring (Fig.~\ref{fig:observable-obl80-Om}b). In the slow rotating end, seasonal variation is weak and wind tend to have a period longer than one year (Fig.~\ref{fig:observable-obl80-Om}c, we verified this multi-year oscillation in a steady-state simulation), which might be similar to the equatorial eddy-mean flow interaction like the equatorial QBO on Earth.

% cloud cover for OBL80-Om
Fig.~\ref{fig:observable-obl80-Om}a shows the cloud cover as a function of latitude and rotation rate. Like the circulation and the wind pattern, the cloud pattern also expands meridionally as rotation rate decreases. In the fast rotating end, the meridional circulation is dominated by adiabatic processes, with an annual-mean upward motion above the equator (Fig.~\ref{fig:regime-varyOm-obl80-ann}g). Supported by this upward motion, the equatorial cloud cover is over 80\% all year round (Fig.~\ref{fig:observable-obl80-Om}e). In the slow rotating end, as the circulation becomes more diabatic-dominant, downward motions take over the equator (Fig.~\ref{fig:regime-varyOm-obl80-ann}d), and the equatorial cloud cover decreases in general. Concurrently, the meridional circulation becomes more seasonal varying (the ratio of the DJF circulation amplitude in Fig.~\ref{fig:regime-varyOm-obl80-djf}a over the annual-mean circulation amplitude in Fig.~\ref{fig:regime-varyOm-obl80-ann}a is larger when rotation is slow), and therefore the cloud cover also shows a stronger seasonal cycle.

\section{Conclusions}
\label{sec:conclusions}

%% what we have done
In this work, we use a state-of-the-art GCM to investigate how the meridional circulation and baroclinic eddies change with insolation and rotation rate, under both high and zero obliquity, and to find potential observable consequences of these changes. We separated the meridional circulation driven by adiabatic processes from that driven by diabatic processes, to be able to study their different characteristics and responses differently to changes in insolation and rotation rate.  

%% main findings
We found a regime transition of the meridional circulations in the high obliquity scenario when changing the insolation or the rotation rate, and the regime transitions have observable signatures in upper atmospheric wind and cloud cover. In the limit of low insolation and fast rotation, the adiabatic processes and the momentum drag dominantly drive the meridional circulation (the adiabatic/momentum dominant regime); while in the opposite limit, the diabatic processes and the thermal forcing dominate (the diabatic/thermal dominant regime). Fixing the rotation rate at 1 times Earth rotation rate, the regime transition occurs when the insolation exceeds 1550 and 1600 W/m$^2$ for the solstice circulation and the annual-mean circulation, respectively, marked by a sudden strengthening of the tropical easterly in the upper atmosphere and a sudden drop of the tropical cloud cover. Fixing the insolation at 1550 W/m$^2$, the transition occurs around 0.7-1 times Earth rotation rate as the rotation rate decreases, accompanied with a similar reduction of tropical cloud cover, but a weakening of the upper atmospheric easterly. The observable signatures of the changes of the insolation and the rotation rate may be distinguishable, because unlike the cloud cover, the upper atmospheric zonal wind is affected by not only the meridional circulation but also the planet's rotation rate. Therefore, observing both of the cloud cover and the upper atmospheric zonal wind may help distinguish high (low) insolation from slow (fast) rotation. Under low obliquity, the diabatic/thermal circulation component always dominates in all of the parameter combinations we explored, consistent with the understanding of the Earth climate \citep{Pfeffer-1987:comparison}.
% The above regime transition of the meridional circulation on high obliquity planets are not evident in \citet{Linsenmeier-Pascale-Lucarini-2015:climate}, probably because the T21 resolution used in their model is not adequate to simulate baroclinic eddies, or because their simulations are constrained close to the outer edge of the habitable zone.

Anther key result is that the dry dynamic core model forced by a reversed meridional temperature gradient in \citet[][KCT]{Kang-Cai-Tziperman-2019:tropical} can explain most of the characteristics of the baroclinic eddies and the momentum-driven meridional circulation in the realistic high-obliquity experiments. The baroclinic eddies under high obliquity are shallow and bottom-amplified, driving a meridional circulation that is also shallow. In the regime that the adiabatic processes dominate, this characteristics are reflected in the total circulation. Therefore, the dry model study (KCT) provides useful insights to the understanding of the high obliquity general circulation, especially the circulation on cool and fast-rotating planets. 

%% caveat
This work proposes potential pathways to observe the circulation regime transition, and more needs to be done to evaluate the phase curves under different situations and compare it with the instrumental accuracy. For simplicity, we here do not discuss or explore the multiple equilibrium states. By gradually increasing the insolation, and reducing the rotation rate, we keep the model in the cold branch. More work is needed to examine whether the two-regime behavior also exists in the other branch. Also, water vapor is the only strong greenhouse gas considered in the radiation scheme used here, and the robustness of our results to atmospheric composition need to be examined.

%% put in perspective
%This work opens up some new questions. E.g., would the Hadley strength and extent follow the same rule through a wide parameter range, given the two-regime behavior? If not, then more theoretical work is needed to get the scaling rule governing each regime, if there is any.

\acknowledgments
The author thanks Prof. Eli Tziperman and Prof. Ming Cai for the insightful and helpful discussion. This work was supported by NASA Habitable Worlds program (grant FP062796-A/NNX16AR85G). We would like to acknowledge high-performance computing support from Cheyenne provided by NCAR's Computational and Information Systems Laboratory, sponsored by the National Science Foundation. The relevant model outputs are archived in \url{https://www.dropbox.com/sh/ugkmaw4kjupx91a/AACXnBEylRoqyx2iqfe9QAH9a?dl=0} and in Harvard Dataverse \url{https://doi.org/10.7910/DVN/XC4E1I}.

%%%%%%%%%%%%%%%%%%% APPENDIX %%%%%%%%%%%%%%%%%%%%
 \appendix
% \section{Appendix information}

%% The reference list follows the main body and any appendices.
%% Use LaTeX's thebibliography environment to mark up your reference list.
%% Note \begin{thebibliography} is followed by an empty set of
%% curly braces.  If you forget this, LaTeX will generate the error
%% "Perhaps a missing \item?".
%%
%% thebibliography produces citations in the text using \bibitem-\cite
%% cross-referencing. Each reference is preceded by a
%% \bibitem command that defines in curly braces the KEY that corresponds
%% to the KEY in the \cite commands (see the first section above).
%% Make sure that you provide a unique KEY for every \bibitem or else the
%% paper will not LaTeX. The square brackets should contain
%% the citation text that LaTeX will insert in
%% place of the \cite commands.

%% We have used macros to produce journal name abbreviations.
%% \aastex provides a number of these for the more frequently-cited journals.
%% See the Author Guide for a list of them.

%% Note that the style of the \bibitem labels (in []) is slightly
%% different from previous examples.  The natbib system solves a host
%% of citation expression problems, but it is necessary to clearly
%% delimit the year from the author name used in the citation.
%% See the natbib documentation for more details and options.
 \renewcommand\thefigure{A\arabic{figure}}  
 \setcounter{figure}{0}
 
\begin{figure*}[!tbh]
 \centering
 \includegraphics[width=\textwidth]{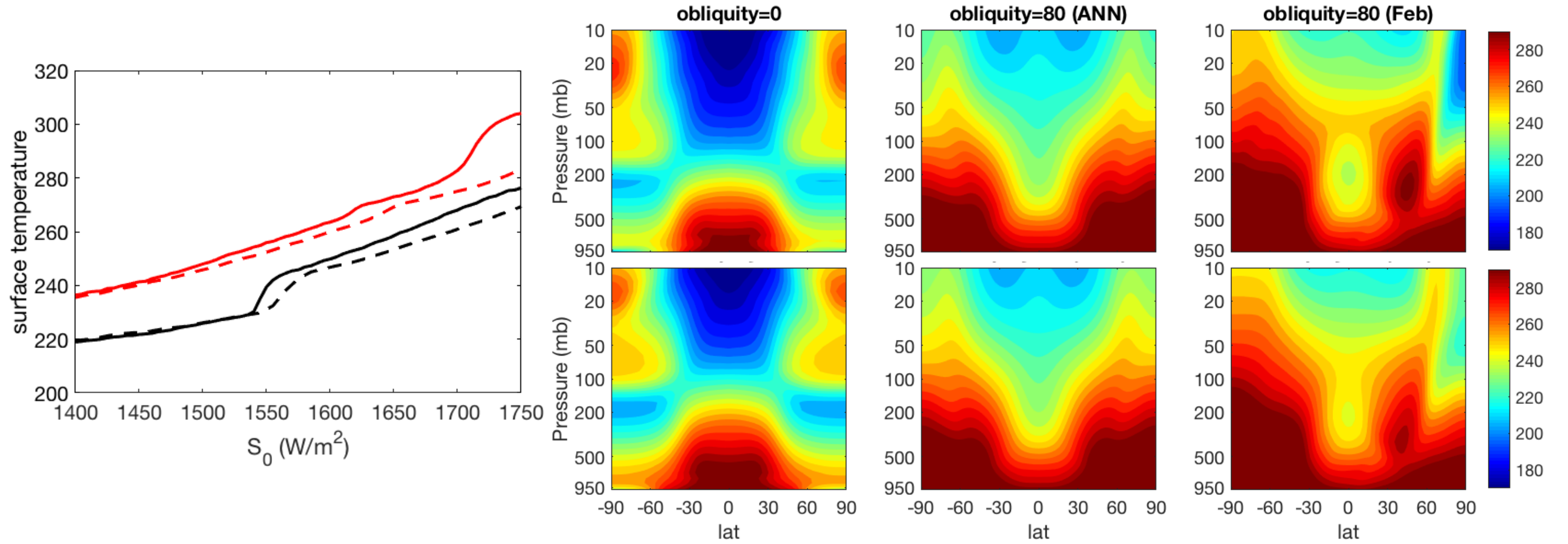}
 \caption{Demonstrations to show that the transient simulations almost reach equilibrium. Panel (a) shows the global-annual mean surface temperature evolution in the slow (default) transient experiments (solid curves), compared with that in the fast transient experiments, where the insolation is tuned up twice as fast (dashed curves). Panel (b-d) show the zonal mean temperature profiles toward the end of the insolation varying simulation (1750~W/m$^2$), for low obliquity annual mean, high obliquity annual mean, and high obliquity February. Panel (e-g) are the same plots for the steady-state experiments, where the insolation is fixed at 1750~W/m$^2$ for 40 years.}
 \label{fig:transient-equilibrium}
\end{figure*}

\begin{figure*}[!tbh]
 \centering
 \includegraphics[width=\textwidth]{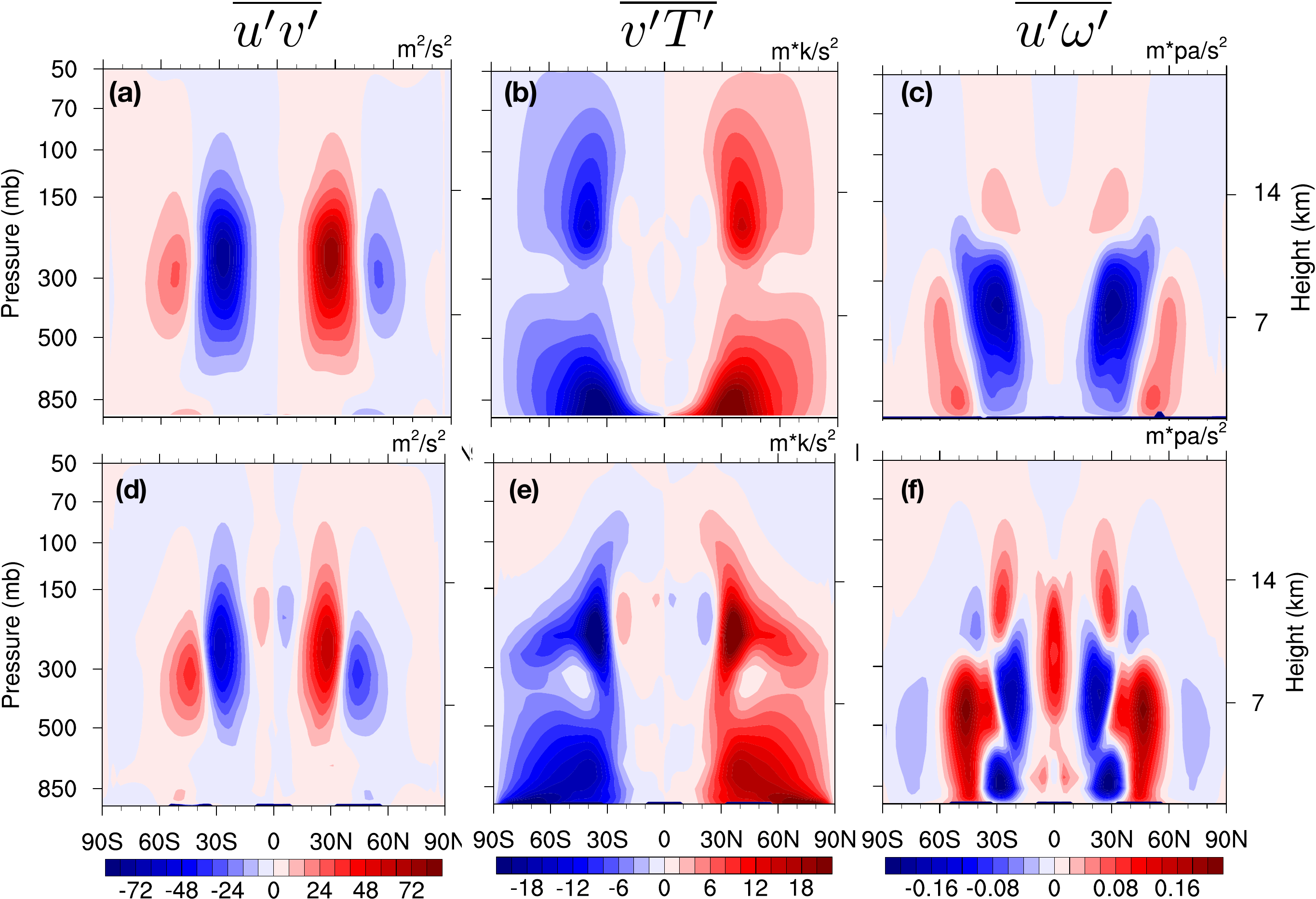}
 \caption{Eddy heat/momentum transports for (a,b,c) the dry dynamic model used in KCT, forced with a normal meridional temperature gradient (equator is warmer than poles), (d,e,f) the 0 degree obliquity experiment forced by $S_0=1360~W/m^2$ insolation. From left to right, shown are meridional eddy momentum transport $\overline{u'v'}$, meridional eddy heat transport $\overline{v'T'}$, and vertical eddy momentum transport $\overline{u'\omega'}$. }
 \label{fig:eddy-transport-obl0}
\end{figure*}

%%%%%%%%%%%%%%%%%%% BIBLIOGRAPHY %%%%%%%%%%%%%%%%%%%%

%% This command is needed to show the entire author+affilation list when
%% the collaboration and author truncation commands are used.  It has to
%% go at the end of the manuscript.
%\allauthors
%\bibliography{export.bib}

\begin{thebibliography}{}
\expandafter\ifx\csname natexlab\endcsname\relax\def\natexlab#1{#1}\fi
\providecommand{\url}[1]{\href{#1}{#1}}
\providecommand{\dodoi}[1]{doi:~\href{http://doi.org/#1}{\nolinkurl{#1}}}
\providecommand{\doeprint}[1]{\href{http://ascl.net/#1}{\nolinkurl{http://ascl.net/#1}}}
\providecommand{\doarXiv}[1]{\href{https://arxiv.org/abs/#1}{\nolinkurl{https://arxiv.org/abs/#1}}}

\bibitem[{Armstrong {et~al.}(2014)Armstrong, Barnes, Domagal-Goldman, Breiner,
  Quinn, \& Meadows}]{Armstrong-Barnes-Domagal-Goldman-et-al-2014:effects}
Armstrong, J.~C., Barnes, R., Domagal-Goldman, S., {et~al.} 2014, Astrobiology,
  14, 277

\bibitem[{Brunini(2006)}]{Brunini-2006:origin}
Brunini, A. 2006, Nature, 440, 1163

\bibitem[{{Carpenter}(1966)}]{Carpenter-1966:study}
{Carpenter}, I.~R.~L. 1966, Astronomical Journal, 71, 142,
  \dodoi{10.1086/109872}

\bibitem[{Chang(1996)}]{Chang-1996:mean}
Chang, E. K.~M. 1996, Journal of Atmospheric Sciences, 53, 113,
  \dodoi{10.1175/1520-0469(1996)053<0113:MMCDBE>2.0.CO;2}

\bibitem[{Correia \& Laskar(2010)}]{Correia-Laskar-2010:tidal}
Correia, A.~C., \& Laskar, J. 2010, Tidal evolution of exoplanets (University
  of Arizona Press)

\bibitem[{Demory {et~al.}(2013)Demory, de~Wit, Lewis, Fortney, Zsom, Seager,
  Knutson, Heng, Madhusudhan, Gillon, Barclay, Desert, Parmentier, \&
  Cowan}]{Demory-Wit-Lewis-et-al-2013:inference}
Demory, B.-O., de~Wit, J., Lewis, N., {et~al.} 2013, arXiv.org,
  \dodoi{10.1088/2041-8205/776/2/L25}

\bibitem[{Ferreira {et~al.}(2014)Ferreira, Marshall, O'Gorman, \&
  Seager}]{Ferreira-Marshall-OGorman-et-al-2014:climate}
Ferreira, D., Marshall, J., O'Gorman, P.~A., \& Seager, S. 2014, Icarus, 243,
  236

\bibitem[{Held(2000)}]{Held-2000:general}
Held, I.~M. 2000, The general circulation of the atmosphere (Woods Hole Lecture
  Notes), 77.
\newblock \url{https://www.whoi.edu/fileserver.do?id=21464&pt=10&p=17332}

\bibitem[{Held \& Hou(1980)}]{Held-Hou-1980}
Held, I.~M., \& Hou, A.~Y. 1980, J. Atmos. Sci., 37, 515

\bibitem[{Held \& Suarez(1994)}]{Held-Suarez-1994:proposal}
Held, I.~M., \& Suarez, M.~J. 1994, Bull. Amer. Meteor. Soc., 75, 1825,
  \dodoi{10.1175/1520-0477(1994)075<1825:APFTIO>2.0.CO;2}

\bibitem[{Huybers \& Wunsch(2005)}]{Huybers-Wunsch-2005:obliquity}
Huybers, P., \& Wunsch, C. 2005, Nature, 434, 491

\bibitem[{Jenkins(2003)}]{Jenkins-2003:gcm}
Jenkins, G.~S. 2003, J. Geophys. Res., 108, \dodoi{10.1029/2001JD001582}

\bibitem[{Kang(2019{\natexlab{a}})}]{Kang-2019:mechanisms}
Kang, W. 2019{\natexlab{a}}, The Astrophysical Journal, 876, L1,
  \dodoi{10.3847/2041-8213/ab18a8}

\bibitem[{Kang(2019{\natexlab{b}})}]{Kang-2019:wetter}
---. 2019{\natexlab{b}}, ApJL

\bibitem[{Kang {et~al.}(2019)Kang, Cai, \&
  Tziperman}]{Kang-Cai-Tziperman-2019:tropical}
Kang, W., Cai, M., \& Tziperman, E. 2019, Icarus,
  \dodoi{https://doi.org/10.1016/j.icarus.2019.04.028}

\bibitem[{Kaspi \& Showman(2015)}]{Kaspi-Showman-2015:atmospheric}
Kaspi, Y., \& Showman, A. 2015, The Astrophysical Journal, 804, 60

\bibitem[{Kasting \& Pollack(1983)}]{Kasting-Pollack-1983:loss}
Kasting, J.~F., \& Pollack, J.~B. 1983, Icarus (ISSN 0019-1035), 53, 479

\bibitem[{Kilic {et~al.}(2018)Kilic, Lunkeit, Raible, \&
  Stocker}]{Kilic-Lunkeit-Raible-et-al-2018:stable}
Kilic, C., Lunkeit, F., Raible, C.~C., \& Stocker, T.~F. 2018, Astrophysical
  Journal, 864, 106

\bibitem[{Kilic {et~al.}(2017)Kilic, Raible, \&
  Stocker}]{Kilic-Raible-Stocker-2017:multiple}
Kilic, C., Raible, C.~C., \& Stocker, T.~F. 2017, Astrophysical Journal, 844,
  147

\bibitem[{Kopparapu {et~al.}(2017)Kopparapu, Wolf, Arney, \&
  et~al.}]{Kopparapu-Wolf-Arney-et-al-2017:habitable}
Kopparapu, R., Wolf, E.~T., Arney, G., \& et~al. 2017, The Astrophysical
  Journal, 845, 5

\bibitem[{Kuo(1956)}]{Kuo-1956:forced}
Kuo, H.-L. 1956, Journal of Atmospheric Sciences, 13, 561

\bibitem[{Laskar \& Robutel(1993)}]{Laskar-Robutel-1993:chaotic}
Laskar, J., \& Robutel, P. 1993, Nature, 361, 608

\bibitem[{Linsenmeier {et~al.}(2015)Linsenmeier, Pascale, \&
  Lucarini}]{Linsenmeier-Pascale-Lucarini-2015:climate}
Linsenmeier, M., Pascale, S., \& Lucarini, V. 2015, Planetary and Space
  Science, 105, 43, \dodoi{10.1016/j.pss.2014.11.003}

\bibitem[{Millholland \& Laughlin(2019)}]{Millholland-Laughlin-2019:obliquity}
Millholland, S., \& Laughlin, G. 2019, Nature Astronomy, 204, 1,
  \dodoi{10.1038/s41550-019-0701-7}

\bibitem[{Neale {et~al.}(2010)Neale, Chen, \&
  Gettelman}]{Neale-Chen-Gettelman-2010:description}
Neale, R.~B., Chen, C.~C., \& Gettelman, A. 2010, NCAR Tech Note

\bibitem[{Paynter \& Ramaswamy(2014)}]{Paynter-Ramaswamy-2014:investigating}
Paynter, D., \& Ramaswamy, V. 2014, JGR, 119, 10,720,
  \dodoi{10.1002/2014JD021881}

\bibitem[{Pfeffer(1987)}]{Pfeffer-1987:comparison}
Pfeffer, R.~L. 1987, Quarterly Journal of the Royal Meteorological Society,
  113, 237

\bibitem[{Rose {et~al.}(2017)Rose, Cronin, \& Bitz}]{Rose-Cronin-Bitz-2017:ice}
Rose, B. E.~J., Cronin, T.~W., \& Bitz, C.~M. 2017, The Astrophysical Journal,
  846, 28

\bibitem[{Rothman {et~al.}(2013)Rothman, Gordon, Babikov, Barbe, Chris~Benner,
  Bernath, Birk, Bizzocchi, Boudon, Brown, Campargue, Chance, Cohen, Coudert,
  Devi, Drouin, Fayt, Flaud, Gamache, Harrison, Hartmann, Hill, Hodges,
  Jacquemart, Jolly, Lamouroux, Le~Roy, Li, Long, Lyulin, Mackie, Massie,
  Mikhailenko, M{\"u}ller, Naumenko, Nikitin, Orphal, Perevalov, Perrin,
  Polovtseva, Richard, Smith, Starikova, Sung, Tashkun, Tennyson, Toon,
  Tyuterev, \& Wagner}]{Rothman-Gordon-Babikov-et-al-2013:hitran2012}
Rothman, L.~S., Gordon, I.~E., Babikov, Y., {et~al.} 2013, Journal of
  Quantitative Spectroscopy and Radiative Transfer, 130, 4

\bibitem[{Schulz \& Zeebe(2006)}]{Schulz-Zeebe-2006:pleistocene}
Schulz, K.~G., \& Zeebe, R.~E. 2006, Earth and Planetary Science Letters, 249,
  326

\bibitem[{Showman {et~al.}(2012)Showman, Fortney, Lewis, \&
  Shabram}]{Showman-Fortney-Lewis-et-al-2012:doppler}
Showman, A.~P., Fortney, J.~J., Lewis, N.~K., \& Shabram, M. 2012, The
  Astrophysical Journal, 762, 24, \dodoi{10.1088/0004-637x/762/1/24}

\bibitem[{Singh \& O'Gorman(2012)}]{Singh-OGorman-2012:upward}
Singh, M.~S., \& O'Gorman, P.~A. 2012, Journal of Climate, 25, 8259

\bibitem[{Tziperman {et~al.}(2006)Tziperman, Raymo, Huybers, \&
  Wunsch}]{Tziperman-Raymo-Huybers-et-al-2006:consequences}
Tziperman, E., Raymo, M., Huybers, P., \& Wunsch, C. 2006, Paleoceanography,
  21, doi:10.1029/2005PA001241

\bibitem[{Wang {et~al.}(2016)Wang, Liu, Tian, Yang, Ding, Zhou, \&
  Hu}]{Wang-Liu-Tian-et-al-2016:effects}
Wang, Y., Liu, Y., Tian, F., {et~al.} 2016, The Astrophysical Journal, 823, L20

\bibitem[{Williams {et~al.}(1998)Williams, Kasting, \&
  Frakes}]{Williams-Kasting-Frakes-1998:low}
Williams, D.~M., Kasting, J.~F., \& Frakes, L.~A. 1998, Nature, 396, 453

\bibitem[{Yang {et~al.}(2014)Yang, Bou{\'e}, Fabrycky, \&
  Abbot}]{Yang-Boue-Fabrycky-et-al-2014:strong}
Yang, J., Bou{\'e}, G., Fabrycky, D.~C., \& Abbot, D.~S. 2014, The
  Astrophysical Journal Letters, 787, L2

\bibitem[{Yang {et~al.}(2016)Yang, Leconte, Wolf, Goldblatt, Feldl, Merlis,
  Wang, Koll, Ding, Forget, \&
  Abbot}]{Yang-Leconte-Wolf-et-al-2016:differences}
Yang, J., Leconte, J., Wolf, E.~T., {et~al.} 2016, The Astrophysical Journal,
  826, 222, \dodoi{10.3847/0004-637x/826/2/222}

\end{thebibliography}

%%%%%%%%%%%%%%%%%%% Tracking change %%%%%%%%%%%%%%%%%%%%
%% Include this line if you are using the \added, \replaced, \deleted
%% commands to see a summary list of all changes at the end of the article.
%\listofchanges

\end{document}